\documentclass[prb,aps,twocolumn,amsmath,amssymb,floatfix,superscriptaddress]{revtex4}

\usepackage{color}
\usepackage{soul}
\usepackage[colorlinks=true, citecolor=blue, urlcolor=blue]{hyperref}
\usepackage{graphics,epstopdf}
\usepackage{epsf,graphics,graphicx}

\definecolor{dred}{rgb}{0.75,0,0}
\usepackage{hyperref}
\usepackage{xcolor}
\usepackage{siunitx}
\usepackage{placeins}        
\usepackage{dblfloatfix}

\begin{document}

\title{New perspective on symmetry breaking in a clean antiferromagnetic chain: Spin-selective transport and NDR phenomenon}

\author{Prabhab Patra}

\affiliation{Physics and Applied Mathematics Unit, Indian Statistical Institute, 203 Barrackpore Trunk Road, Kolkata-700 108, India}

\author{Santanu K. Maiti}

\email{santanu.maiti@isical.ac.in}

\affiliation{Physics and Applied Mathematics Unit, Indian Statistical Institute, 203 Barrackpore Trunk Road, Kolkata-700 108, India}

\begin{abstract}
	
The primary requirement for achieving spin-selective electron transfer in a nanojunction possessing a magnetic system with zero net
magnetization is to break the symmetry between the up and down spin sub-Hamiltonians. Circumventing the available approaches, in the 
present work, we put forward a new mechanism for symmetry breaking by introducing a bias drop along the functional element. To demonstrate 
this, we consider a clean magnetic chain with antiparallel alignment of neighboring magnetic moments. The junction is modeled within a 
tight-binding framework, and spin-dependent transmission probabilities are evaluated using wave-guide theory. The corresponding current
components are obtained through the Landauer-B\"{u}ttiker formalism. Selective spin currents, exhibiting a high degree of spin polarization, 
are obtained over a wide bias region. Moreover, the bias-dependent transmission profile exhibits negative differential resistance 
(NDR), another important aspect of our study. We examine the results under three different potential profiles, one linear and two 
non-linear, and in each case, we observe a favorable response. This work may offer a new route for designing efficient spintronic devices 
based on bias-controlled magnetic systems with vanishing net magnetization.    

\end{abstract}

\maketitle

\section{Introduction}

Spintronics~\cite{stronics,ami1,ami2,ami3} has emerged as a major field of study in modern electronics. The central idea of this 
field revolves 
around the manipulation and control of the spin degrees of freedom. To utilize spin for the fabrication of advanced, compact, smart, 
and powerful electronic devices, such as spin filters~\cite{d1,d1n1,d1n2}, spin diodes~\cite{d2}, spin transistors~\cite{d3}, 
memory devices\cite{d4,d5}, and many more~\cite{d1n3}, it is essential to 
create an imbalance between up and down spin electrons. That is why, one of the primary challenges in building such devices is the separation 
of spin energy channels to achieve spin-selective transmission. For decades, the common approach has involved using ferromagnetic (FM)~\cite{f1,f2,f3} materials as spin-selective functional elements. However, 
due to the large resistivity mismatch at junction interfaces~\cite{r1,r2}, this approach has seen 
quite limited success. As a result, attention has shifted toward systems with spin-orbit (SO) coupling~\cite{so1,so2,so3,so4}. The SO coupling, on the other hand, is typically 
too weak compared to the electronic hopping strength~\cite{ht}, which hinders its effectiveness. To overcome these 
limitations, researchers have increasingly turned to magnetic systems possessing zero net magnetization, more specifically we can say antiferromagnetic systems~\cite{af1,af2,af3,af4} 
as functional elements due to their faster operational speed, ability to function at higher frequencies, and absence of stray magnetic fields.

For a {\em clean} (disorder-free) magnetic system with vanishing net magnetization, in presence of uniform electron 
hopping, the sub-Hamiltonians $H_{\uparrow}$ and $H_{\downarrow}$, associated
with up and down spin electrons, are {\em symmetric} to each other, which makes it difficult to establish a mismatch between up and down 
spin energy channels. To break the symmetry between $H_{\uparrow}$ and $H_{\downarrow}$, some proposals have been put forward, such as
introducing substitutional disorder into the system, incorporating hopping asymmetry in different segments, or applying a transverse 
electric field~\cite{f3,sy1}. In the present article, we propose a new prescription for symmetry breaking by considering a bias drop along the 
functional element that bridges contact electrodes. In most studies of nanoscale junctions, the applied bias is assumed to drop entirely 
at the interfaces between the conductor and the electrodes. This simplification is often reasonable for too short conductors. Incorporating 
a bias drop along the conductor itself provides a more realistic description of electron transport and can qualitatively alter the 
transport characteristics.
\begin{figure}[ht]
{\centering \resizebox*{7.5cm}{1.5cm}{\hskip 1cm \includegraphics{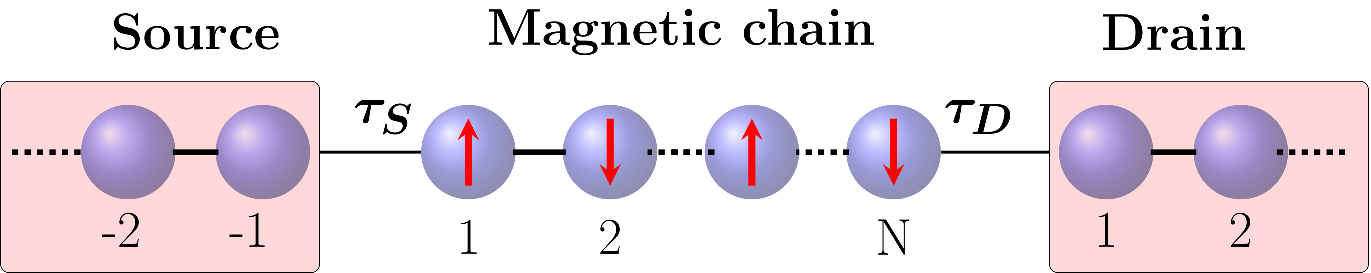}}\par}           
\caption{(Color online). Schematic of the junction setup where a one-dimensional antiferromagnetic chain with zero net magnetization is 
clamped between two nonmagnetic 1D electrodes, namely, source (S) and drain (D). The red arrows are the magnetic moments directed along 
the $+Z$ and $-Z$ directions alternatively.}
\label{fig1}
\end{figure}
To substantiate these facts, we analyze a magnetic nanojunction where an antiferromagnetic (AFM)~\cite{afnw1,afnw2,afnw3,afnw4,afnw5,afnw6} 
chain is coupled to source and drain electrodes (Fig.~\ref{fig1}). 
The neighboring magnetic moments in the chain are arranged in an antiparallel configuration. These localized magnetic moments scatter 
itinerant electron spins via the usual spin-moment exchange interaction, giving rise to spin-dependent transport phenomena.

Illustrating the nanojunction within a tight-binding (TB) framework~\cite{TB1,TB2,TB3}, we compute the spin-dependent transmission probabilities following 
the wave-guide theory~\cite{WG1,WG2,WG3,WG4}, and evaluate the current components using the Landauer-B\"{u}ttiker~\cite{LB1} prescription. In presence of a potential 
drop along the AFM chain, a mismatch occurs between the up and down spin currents, and the degree of mismatch is measured by evaluating
spin polarization coefficient~\cite{SP1}. Our results provide a high degree of spin polarization for a broad range of bias voltage. By inspecting 
the junction current, we also observe the emergence of negative differential resistance (NDR) effect~\cite{ndr1,ndr2,ndr3,ndr4,ndr5,ndr6}, a phenomenon where the current decreases with increasing applied voltage after a certain threshold. This effect was first discovered by Leo Esaki\cite{LE} in tunnel diodes and has become a cornerstone in modern electronics. The NDR is essential for developing self-switching circuits\cite{Nd1}, amplifiers\cite{Nd2}, memory circuits\cite{Nd3}, and more\cite{Nd3,Nd4}. A relaxation oscillator\cite{RO} using a tunnel diode utilizes its negative differential resistance region to achieve self-sustained oscillators. The interplay between NDR behavior and a large inductance induces a periodic transition across the current-voltage characteristic, supporting high-frequency signal generation without external switching elements.  
The extent of NDR is typically quantified by the peak-to-valley current ratio (PVCR)~\cite{ndr5,PVCR1,PVCR2}. 
A larger PVCR indicates more pronounced NDR behavior, and in our work, we put emphasis to achieve it. 
Additionally, we investigate how the threshold voltage~\cite{vth1,vth2} $V_{TH}$, the voltage corresponding to 
the peak current, changes with system temperature, spin-dependent scattering parameter, and the chain-to-electrode coupling strength. At
the end, we also consider the effect of disorder, to make the model more realistic. 

We analyze the results for three distinct potential profiles~\cite{prof1}, one linear and two nonlinear, and find favorable responses in all scenarios. 
Our findings may suggest a promising pathway toward the design of efficient spintronic devices driven by bias-controlled magnetic systems 
with zero net magnetization.

The rest of the work is organized as follows.
Section II presents the theoretical formulation, including the junction setup, Hamiltonian, and relevant calculations. Section III discusses the numerical results in detail, experimental perspectives, and possible design strategies for realizing such a system in the laboratory. Finally, Section IV summarizes the key findings and an outlook.

\section{Model and Theoretical Framework}

This section illustrates the junction setup, TB Hamiltonian of the full system, and the required theoretical steps for calculating the 
results of our study.

\subsection{Junction setup and the Hamiltonian}

We start by describing the junction setup shown in Fig.~\ref{fig1}, where the central quantum system is a one-dimensional (1D) TB 
AFM chain consisting of $N$ (even) lattice sites. Each site contains a local magnetic moment, and the neighboring magnetic moments 
are aligned alternatively along the $+Z$ and $-Z$ directions. The AFM chain is clamped between two 1D, perfect, non-magnetic electrodes, 
source (S) and drain (D).

The Hamiltonian of the full junction setup can be written as,
\begin{equation}
\mathbb{H}=H_{AFM}+H_S+H_D+H_{cpl}
\end{equation}
where the first three sub-Hamiltonians are associated with the AFM chain, source, and the drain electrodes, and 
\begin{figure}[ht]
	{\centering \resizebox*{8cm}{4.75cm}{\hskip 1cm \includegraphics{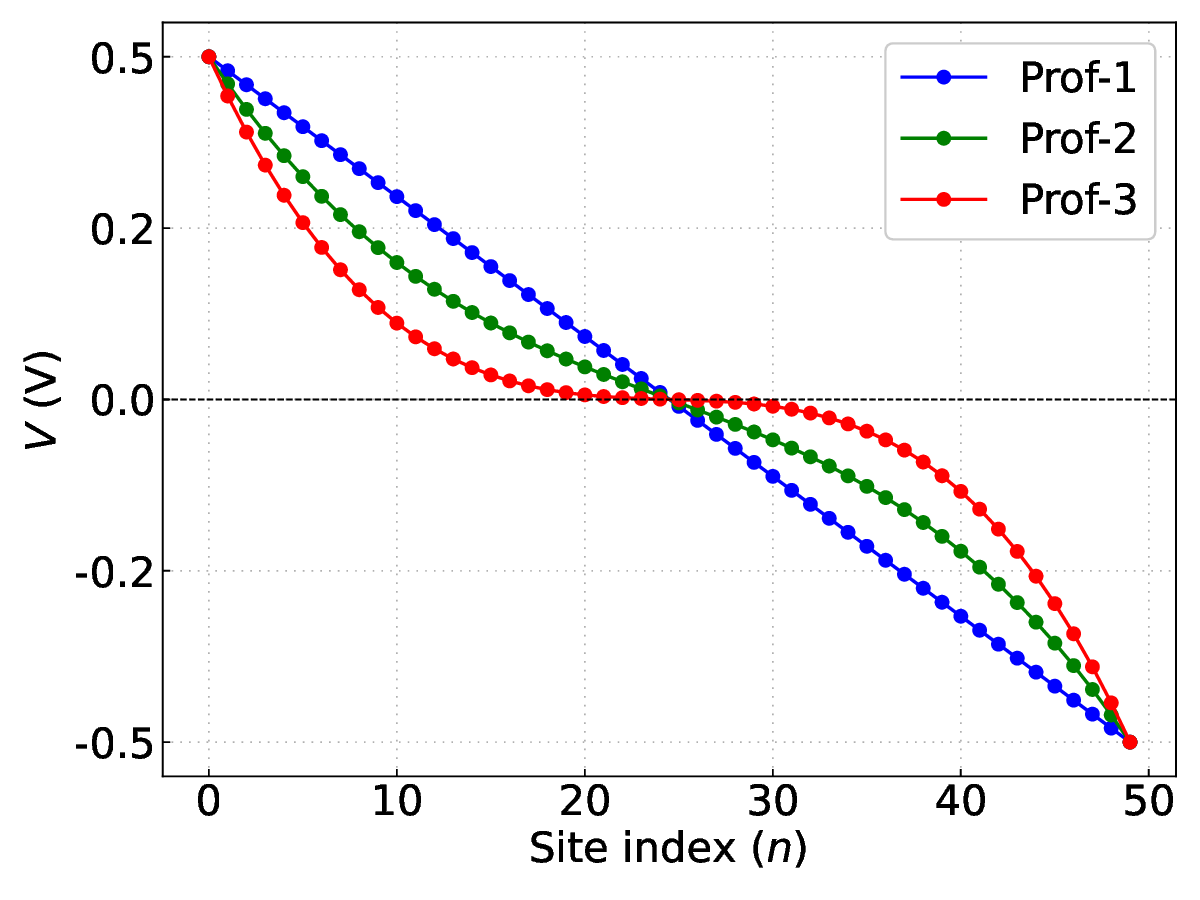}}\par}
	\caption{(Color online). Bias drop along the AFM chain as a function of site index. Three different potential profiles are shown, where one is linear and the other two are nonlinear.}
	\label{fig:f2}
\end{figure}
the last one is involved with the coupling of the AFM chain and the contact electrodes. All these sub-Hamiltonians are expressed within 
a TB framework and their explicit forms are given below. The sub-Hamiltonian $H_{AFM}$ reads as~\cite{TB1,TB2,TB3}
\begin{equation}
H_{AFM}=\sum_n \mathbf{c_n^{\dagger}}\left(\mathbf{\epsilon_n^{eff}}-\vec{\mathbf{h}}_\mathbf{n}.\boldsymbol{\vec{\sigma}}\right)\mathbf{c_n}
+\sum_n(\mathbf{c_{n+1}^{\dagger}}\mathbf{t}\mathbf{c_n}+h.c.)
\end{equation}
where $\mathbf{c_n^{\dagger}}=\begin{pmatrix}c_{n\uparrow}^{\dagger} & c_{n\downarrow}^{\dagger}\end{pmatrix}$. 
$c_{n\sigma}^{\dagger}$ ($c_{n\sigma}$) is the creation (annihilation) operator at site `$n$' for an electron with spin 
$\sigma$ $(\sigma= \uparrow,\downarrow)$. $\mathbf{t}$ is a $(2\times 2)$ diagonal matrix, where the diagonal elements represent the nearest-neighbor hopping (NNH) strength of up and down spin electrons. Both hopping strengths are equal, and denoted by the 
parameter $t$.           
$\left(\mathbf{\epsilon_n^{eff}}-\vec{\mathbf{h}}_\mathbf{n}\cdot \boldsymbol{\vec{\sigma}}\right)$ is a $(2 \times 2)$ site energy matrix, where $\vec{\mathbf{h}}_\mathbf{n}\cdot \boldsymbol{\vec{\sigma}}$ term is responsible for spin-dependent scattering. $\vec{\mathbf{h}}_\mathbf{n}\cdot \boldsymbol{\vec{\sigma}}=J\langle \boldsymbol{\vec{S}_n} \rangle \cdot \boldsymbol{\vec{\sigma}}$, 
where $\langle \boldsymbol{\vec{S}_n} \rangle$ is the net localized spin at site $n$, $\boldsymbol{\vec{\sigma}} \,\{\boldsymbol{\sigma_x,\sigma_y,\sigma_z } \}$ is the Pauli spin vector, and $J$ is the exchange coupling strength. Any arbitrary 
orientation of $\langle\boldsymbol{\vec{S}_n}\rangle$, and hence, $\boldsymbol{\vec{h}_n}$ can be described by the usual polar angle 
$\theta_n$ and azimuthal angle $\phi_n$ in the spherical polar coordinate system. In our chosen magnetic system, the magnetic moments 
are oriented in $\pm Z$ directions (spin quantization axes), and therefore, 
$\boldsymbol{\vec{h}_n} \cdot \boldsymbol{\vec{\sigma}}=\text{diag}(h_n, -h_n)$. $h_n$ is the magnitude, and is commonly referred to as 
spin-dependent scattering factor. $\mathbf{\epsilon_n^{eff}}=\text{diag}(\epsilon_n^{\small \mbox{eff}}, 
\epsilon_n^{\small \mbox{eff}})$ 
is the site energy matrix in the absence of spin-moment coupling, where $\epsilon_n^{\small\mbox{eff}}$ is written as a sum
$\epsilon_n^{\small\mbox{eff}}=\epsilon_n^{V=0}+\epsilon_n(V)$. $\epsilon_n^{V=0}$ denotes the site energy for the zero-bias 
$(V=0)$ condition. In the presence of a non-zero $V$, when the bias drop occurs along the AFM, the site energies are voltage dependent 
(denoted by the term $\epsilon_n(V)$). We consider three distinct potential profiles along the chain (Fig.~\ref{fig:f2}), among 
which one is linear and the other two are non-linear. We classify the profiles as prof-1, 2, and 3, respectively. In each case, the drop 
is considered symmetrically across the center of the chain. At site $1$, the potential is $V/2$, while at site $N$ it is $-V/2$.
The bias-dependent site energies $\epsilon_n(V)$ are chosen following the distributions shown in Fig.~\ref{fig:f2}, and their 
explicit functional forms are as follows.
\vskip 0.02cm
\noindent 
$\bullet$ Prof-1: $\epsilon_n(V)=\frac{eV}{2}-\frac{neV}{(N+1)}$, \\
$\bullet$ Prof-2: $\epsilon_n(V)=-\frac{eV}{2} \left(\frac{2n-N-1}{N-1}\right) \left[(1-\lambda) \right. \\
\left. \hskip 2.8cm +\lambda\left(\frac{2n-N-1}{N-1}\right)^2\right]$ with $\lambda=0.5$, \\
$\bullet$ Prof-3: The same expression as mentioned in prof-2, with $\lambda=0.96$.
\vskip 0.01cm
\noindent
Here, $e$ is the electronic charge, and $\lambda$ is a parameter that controls the flatness of the profile. In this context, it is relevant 
to point out that finding an exact potential distribution along the chain within a tight-binding framework is a difficult task, as it 
involves many-body calculations. In this work, we consider both linear and non-linear potential distributions, in accordance with previous
theoretical studies in which the bias drop along the system has been taken into account, and also considering situations corresponding 
to different physical systems. For a linear bias drop, it is straightforward to write down the expression for $\epsilon_n(V)$. In contrast, 
for a non-linear profile, the expression mentioned above is not unique. One may also consider other similar expressions that satisfy the 
trend shown in Fig.~\ref{fig:f2}.

The source and drain electrodes are assumed to be perfect, one-dimensional, and non-magnetic, and their Hamiltonians can be expressed 
in a general form as,  
\begin{equation}
H_{S(D)}=\sum_{n}\mathbf{d}_n^{\dagger}\mathbf{\epsilon_0}\mathbf{d}_n+\sum_{n}(\mathbf{d}_{n+1}^{\dagger}\mathbf{t_0}\mathbf{d}_{n}+h.c.)
\end{equation}
where $\mathbf{d}$ will be replaced by $\mathbf{a}$ for the source and $\mathbf{b}$ for the drain, to distinguish them clearly. The matrices $\mathbf{\epsilon_0}$ and $\mathbf{t_0}$ are taken as $\epsilon_0\mathbf{I}$ and $t_0\mathbf{I}$ respectively, where $\mathbf{I}$ is a $(2\times 2)$ identity matrix. Here $\epsilon_0$ and $t_0$ denote the site energy and nearest-neighbor hopping strength, respectively, in the electrodes.

The sub-Hamiltonian $H_{cpl}$, describes the coupling between the AFM chain and the contact electrodes, is written as 
\begin{equation}
H_{cpl}=\mathbf{c}_1^{\dagger}\mathbf{\tau_S}\mathbf{a}_{-1}+\mathbf{c}_N^{\dagger}\mathbf{\tau_D} \mathbf{b}_1+h.c.
\end{equation}
where $\tau_S$  and $\tau_D$ are the coupling strengths
 
\subsection{Transmission probability} 

Computing the transmission probability is essential for finding the junction current. Several methods are available in the literature to obtain 
the transmission profile. In the present work, we utilize wave-guide theory, a standard tool~\cite{WG1,WG2,WG3,WG4}, in which a set of 
coupled equations containing wave amplitudes at different lattice sites are solved.

For our AFM system, since the magnetic moments are aligned along the spin-quantized directions, the Hamiltonian $H_{AFM}$ can be 
written as a sum of two sub-Hamiltonians, $H_{\uparrow}$, $H_{\downarrow}$, corresponding to spin-up and spin-down components i.e., 
$H_{AFM}=H_{\uparrow}+H_{\downarrow}$. Under this situation, the transmission probability can be computed separately for each 
spin channel, and hence, the spin index is omitted from the coupled equation presented below, without any loss of generality.

The coupled equations involving wave amplitudes at different lattice sites, for any spin species, are written as,
\begin{subequations}
\begin{flalign}
    & (E - \mathcal{E}_n) C_n = t C_{n+1} + t C_{n-1}, \ \forall \ n \in \ 2, 3, \ldots, N - 1\ \label{eq:7a} \\
    & (E - \mathcal{E}_n) C_1 = t C_2 + \tau_S A_{-1} \label{eq:7b} \\
    & (E - \mathcal{E}_n) C_N = t C_{N-1} + \tau_D B_{1} \label{eq:7c} \\
    & (E - \epsilon_0) A_{-1} = \tau_S C_1 + t_0 A_{-2} \label{eq:7d} \\
    & (E - \epsilon_0) B_1 = \tau_D C_N + t_0 B_2. \label{eq:7e}
\end{flalign}
\end{subequations}
These coupled equations are obtained from the wave function of the junction, given by,
\begin{equation}
|\psi\rangle=\left[ \sum_p A_p \mathbf{a}_p^{\dagger}+\sum_q B_q\mathbf{b}_q^{\dagger}+\sum_n C_n\mathbf{c}_n^{\dagger}\right] |0\rangle
\end{equation}
where $A_p$, $B_q$, and  $C_n$ are the amplitudes of electrons at site $p$, $q$, and $n$, corresponding to the source, drain, and the 
AFM chain, respectively. Depending on the spin species, the site energy $\mathcal{E}_n$ is modified. For the spin-up case, 
$\mathcal{E}_n=\epsilon_n^{\small\mbox{eff}}-h_n$, whereas for the spin-down case, $\mathcal{E}_n=\epsilon_n^{\small\mbox{eff}}+h_n$.
The amplitudes $A_p$ and $B_q$, associated with the electrodes, are expressed as,
\begin{subequations}
\begin{flalign}
    & A_p=e^{ik(p+1)}+re^{-ik(p+1)} \label{eq:9a} \\
    & B_q=se^{ikq}. \label{eq:9b}
\end{flalign}
\end{subequations}
Here, the variables $r$ and $s$ denote the reflection and transmission amplitudes, respectively. We assume a plane-wave incidence of 
unit amplitude from the source end. The factor $k$ denotes the wave number and is related to the energy $E$ through the dispersion 
relation of the $1$D electrode $E=\epsilon_0+2t_0\cos(k)$. In the expression of $A_p$, there are two terms: one corresponding to 
the incident wave and the 
other to the reflected wave. In contrast, $B_q$ contains only the transmitted component, as there is no reflection in the drain electrode. 
To solve the coupled equations, we need to express $A_{-2}$ and $B_2$ in Eqs.~\ref{eq:7d} and \ref{eq:7e} respectively, in terms of 
$A_{-1}$ and $B_1$, and they are: $A_{-2}=A_{-1} e^{ik}-2 i \sin(k)$ and $B_2=B_1 e^{ik}$.

Using the expressions of $A_{-2}$ and $B_2$, we solve the ($N+2$) coupled equations (Eqs.~\ref{eq:7a}-\ref{eq:7e}) to 
determine the ($N+2$) variables $C_1,C_2.......C_N$, $A_{-1}$, and $B_1$. We do this exercise for two different spin sub-spaces and then 
taking $\left |B_1\right|^2$, we get up and down spin transmission probabilities $T_{\uparrow}$ and $T_{\downarrow}$.

It should be noted that the transmission probabilities acquire an explicit dependence on the applied voltage due to the bias drop along 
the chain, and thus, $T_{\uparrow}$ and $T_{\downarrow}$ become functions of both the energy $E$ and the bias voltage $V$.

\subsection{Junction current and NDR }
 
The junction current is calculated using transmission probability, following the Landauer-B\"{u}ttiker formalism~\cite{LB1}, 
which is defined as,
\begin{equation}
I(V)=\frac{e}{2\pi \hbar}\int T(E,V)(f_S-f_D)\,dE
\end{equation}
where $e$ is the electronic charge, $\hbar$ is the reduced Planck's constant, and $f_S$ and $f_D$ are the 
Fermi-Dirac distributions for the  source and drain, respectively.
These functions are written as: $f_{S(D)}=1/\left(1+\exp[(E-\mu_{S(D)})/k_B\Theta]\right)$, where $k_B$ is the Boltzmann constant and 
$\Theta$ is the equilibrium temperature. $\mu_S$ and $\mu_D$ are the electrochemical potentials of S and D, respectively, and in presence
of bias $V$, these are: $\mu_S=E_F+eV/2$ and $\mu_D=E_F-eV/2$, where $E_F$ represents the equilibrium Fermi energy.     

As the transmission probability is voltage dependent, we find a possibility to obtain the NDR phenomenon, which can be visualized from 
our results in the forthcoming sub-section. To quantify the strength of NDR, we use the peak-to-valley 
ratio~\cite{PVCR4}, and it is given by,
\begin{equation}
 PVCR=\frac{I_{peak}}{I_{valley}}
\end{equation}
where $I_{peak}$ and $I_{valley}$ are maximum and minimum currents in the NDR regions, respectively. Higher ratio leads to a strong NDR phenomenon.

\subsection{Spin polarization}
 
The other quantity of our interest, viz, spin polarization coefficient\cite{SP1,SS,SG}, is defined as, 
\begin{equation}
SP=\frac{I_{\uparrow}-I_{\downarrow}}{I_{\uparrow}+I_{\downarrow}}\times 100 \%,
\end{equation}
where $I_{\uparrow}$ and $I_{\downarrow}$ are the spin-resolved current components. When any component drops to zero, a hundred percent $SP$ is obtained. For other cases, we find intermediate values. Our aim is to find a large SP, as much as it is possible.

\section{Numerical Results and Discussion}

This section includes and analyzes all the essential results. Our central focus is to inspect the critical role played 
\begin{table}[ht]
\caption{Physical parameters that are unchanged throughout the calculations.}
\vskip 0.15cm
\begin{tabular}{|l|c|}
		\hline \hline
		{\bf Physical Parameter} & {\bf Parameter value}   \\ \hline
		On-site energy of electrodes: $\epsilon_0$ & $0\,$eV  \\ \hline
		NNH strength of electrodes: $t_0$ & $3\,$eV  \\\hline 
		On-site energy of clean AFM: {\tiny $\epsilon_n^{V=0}$} & $0\,$eV \\ \hline 
		NNH strength of AFM: $t$ & $1\,$eV   \\ \hline \hline
	\end{tabular}
	\label{tab1}
\end{table}
by the potential drop along the clean AFM system on spin dependent transport phenomena. Three different types of potential profiles 
are taken into account (Fig.~\ref{fig:f2}), and their effects will be discussed in detail. Before delving into the 
results, we specify in Table~\ref{tab1} the parameter values that remain unchanged throughout the calculations. 
For a given set of plots, some parameters are kept constant, and unless stated otherwise, these parameters are as follows: the 
AFM chain-to-electrode coupling strengths $\tau_S = \tau_D = 1\,$eV, the spin-dependent scattering strength $h_n=h= 0.5\,$eV for all 
$n$, the equilibrium temperature $\Theta = 0$, and the number of atomic sites in the AFM chain $N=20$.
\begin{figure}[ht]
	\centering
	\hspace*{-3.3cm}
	\begin{minipage}{0.48\columnwidth}
		\centering
		\resizebox{7.5cm}{5cm}{\includegraphics{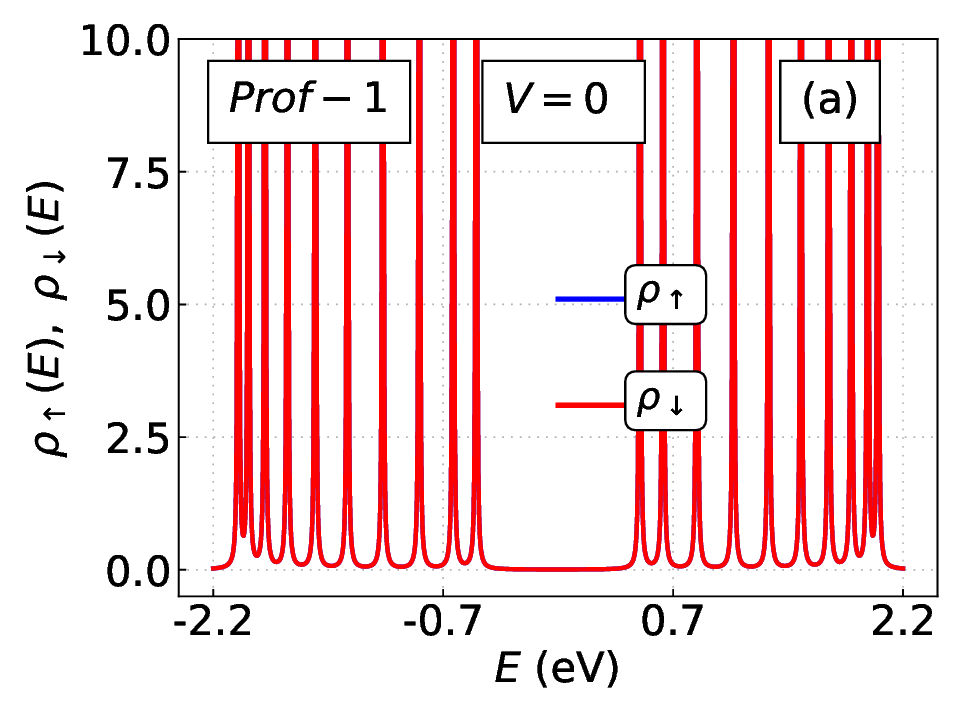}}
	\end{minipage}

   \hspace*{-3.3cm}
	\begin{minipage}{0.48\columnwidth}
		\centering
		\resizebox{7.5cm}{5cm}{\includegraphics{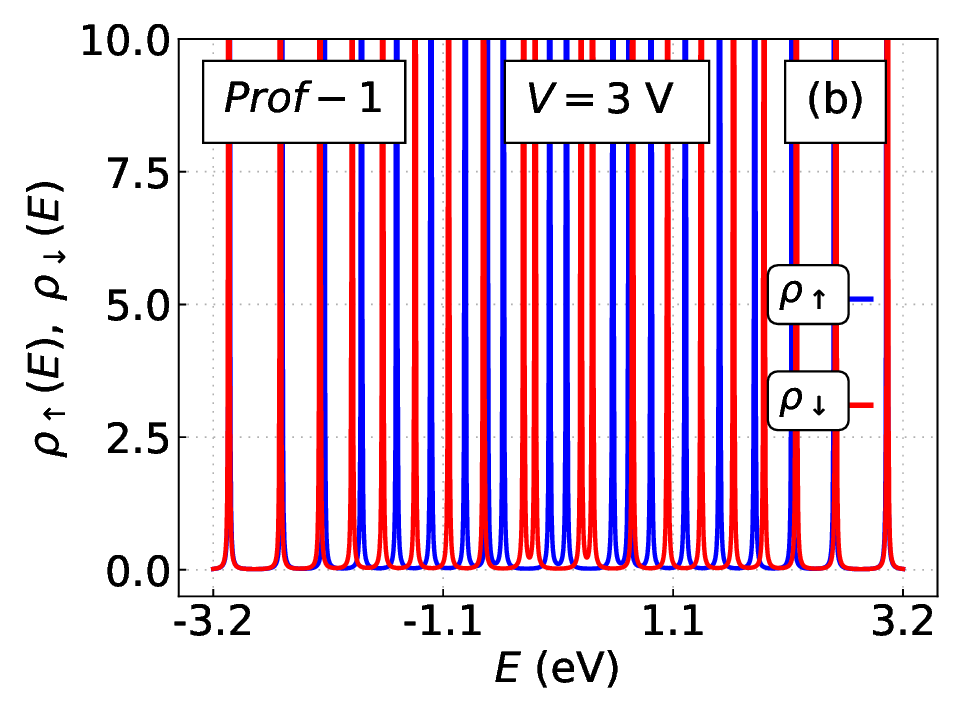}}
	\end{minipage}     
\caption{ (Color online). Spin-specific density of states (blue $ \rightarrow $ up spin, red $\rightarrow$ down spin) as a function of 
energy at two biased conditions for the clean AFM system where (a) $V=0$ and (b) $V=3\,$V. The bias drop is considered following 
prof-1, as illustrated in 
		Fig.~\ref{fig:f2}.}
	\label{fig:f3}
\end{figure}
The remaining parameters are specified in the relevant parts of the discussion. All other energies are measured in eV.

Let us begin with spin specific density of states, which always gives the simplest level of description of the available energy channels for electron transfer between the reservoirs.
The results are shown in Fig.~\ref{fig:f3} for two different conditions: (a) when the bias drop along the chain is zero, and (b) in the presence of a finite bias drop through the chain. The linear bias drop (prof-1) is considered. The results are undoubtedly interesting and important. For the case of zero bias drop, up and down spin DOS spectra are exactly identical viz, the red 
and blue lines completely overlap with each other (Fig.~\ref{fig:f3}(a)).
\begin{figure}[ht]
	\centering
	\hspace*{-3.3cm}
	\begin{minipage}{0.42\columnwidth}
		\centering
		\resizebox{7.5cm}{5cm}{\includegraphics{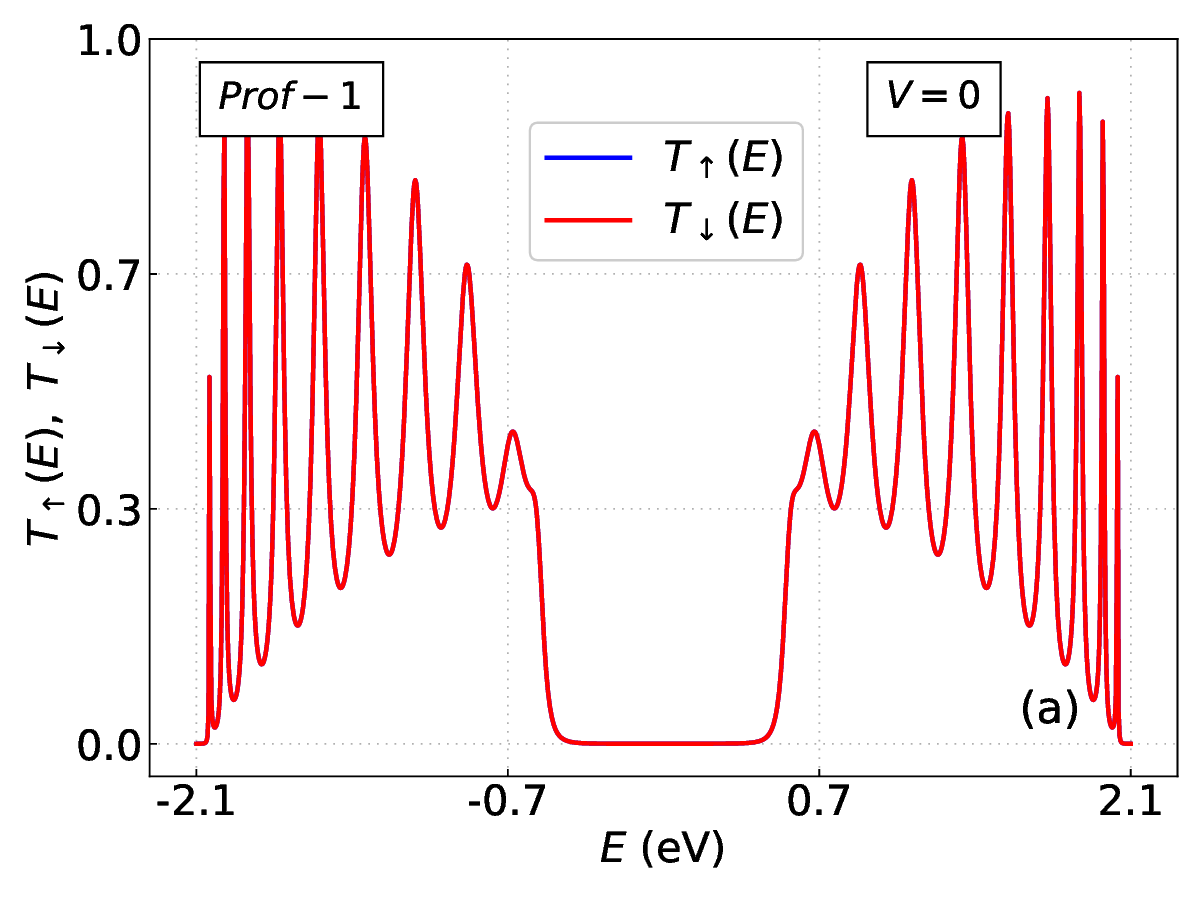}}
	\end{minipage}

	\hspace*{-3.3cm}
	\begin{minipage}{0.42\columnwidth}
		\centering
		\resizebox{7.5cm}{5cm}{\includegraphics{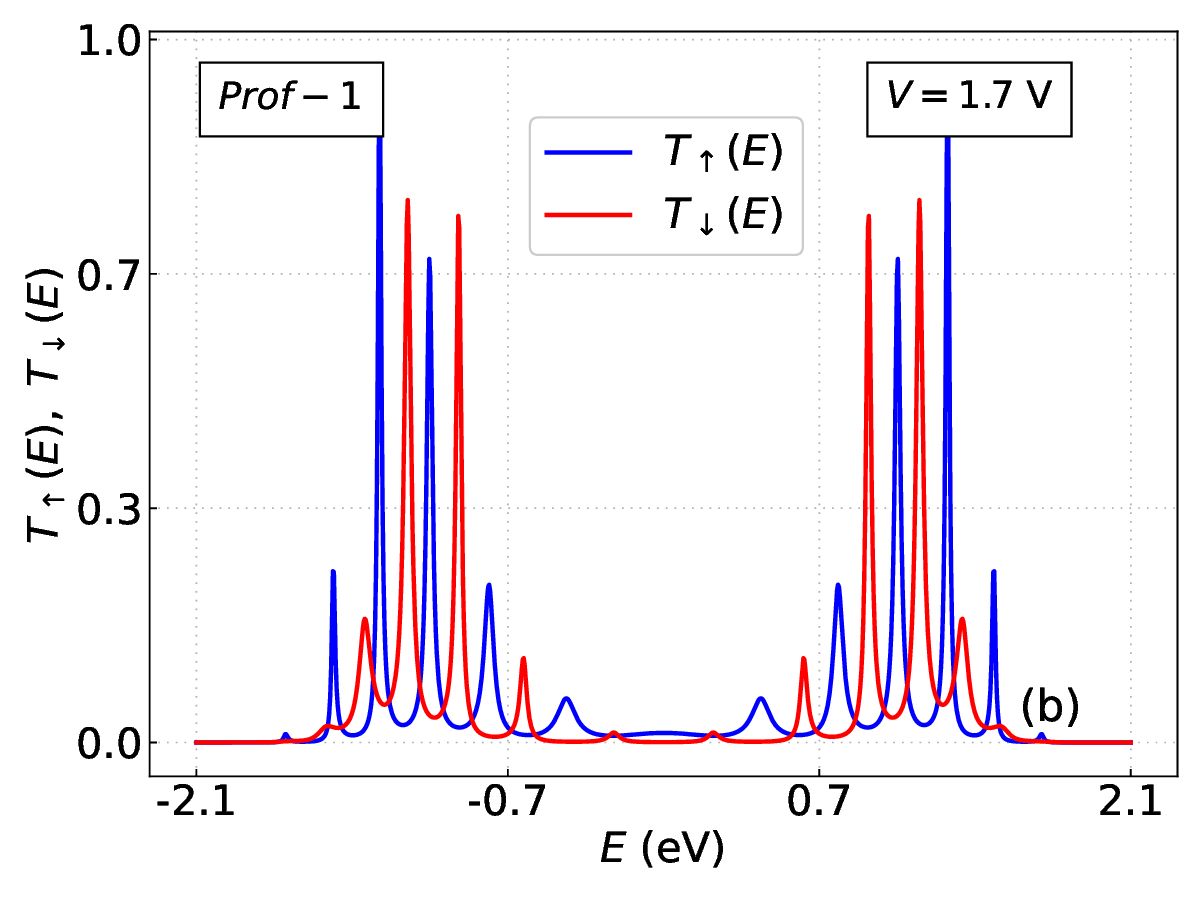}}
	\end{minipage}  

	\hspace*{-3.3cm}
	\begin{minipage}{0.42\columnwidth}
		\centering
		\resizebox{7.5cm}{5cm}{\includegraphics{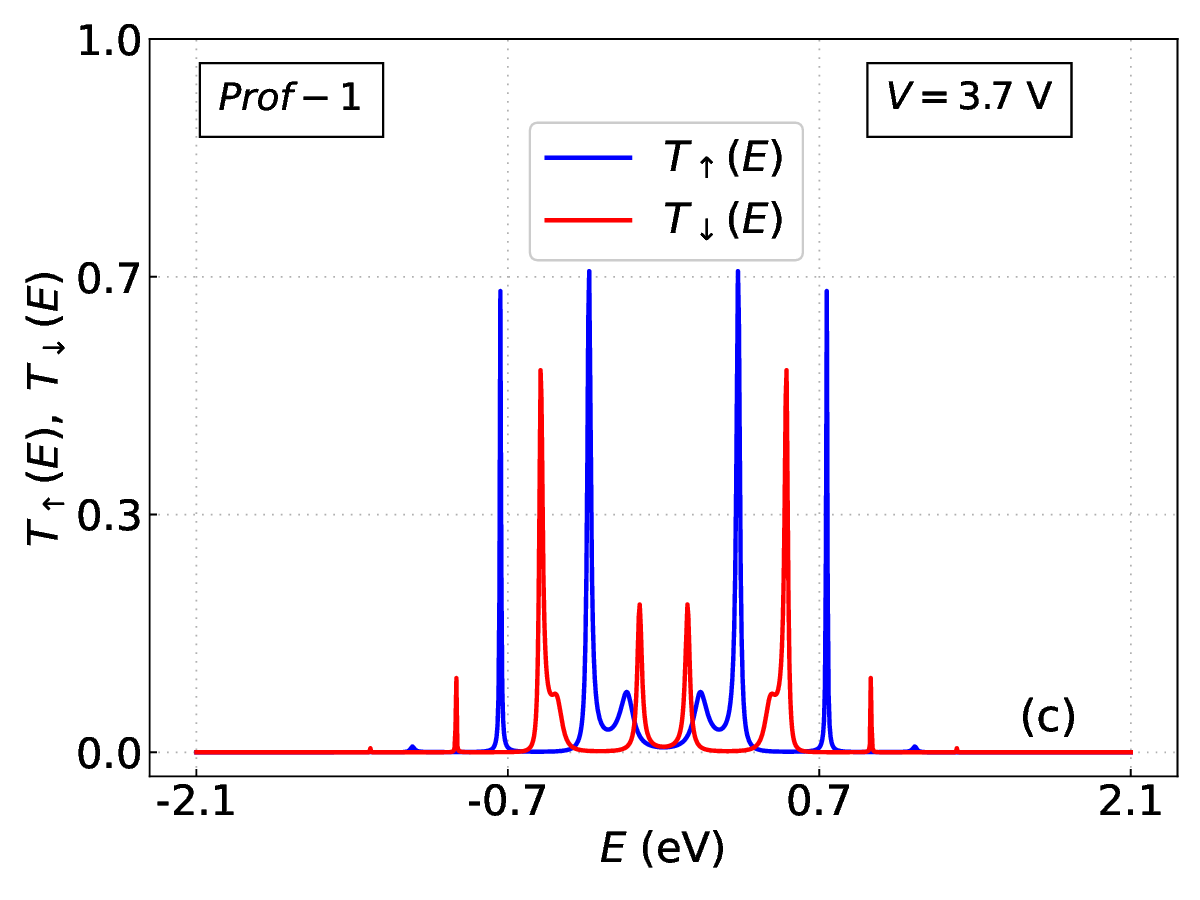}}
	\end{minipage}
	\caption{(Color online). Up (blue) and down (red) spin transmission probabilities as a function of energy at three distinct biased conditions for the clean AFM chain, where (a) $V=0$, (b) $V=1.7\,$V, and (c) $V=3.7\,$V. The bias drop is the same as taken in Fig.~\ref{fig:f2}.}
	\label{fig:f4}
\end{figure}
It clearly suggests that $H_{\uparrow}$ and $H_{\downarrow}$ are symmetric to each other. The symmetric nature can be understood as follows. 
For the clean AFM chain, the site energies for up spin electrons at different sites are: $-h,h,-h,h,.....$, and for the down spin case, they are: $h,-h,h,-h,.....$. Since the hopping strengths are identical, $H_{\uparrow} $ and $H_{\downarrow}$ lead to the same set of eigenenergies, resulting in identical DOS. Once the bias drop along the chain is incorporated, the effective site energies get modified distinctly for the two spin cases, and hence, the symmetry between $H_{\uparrow}$ and $H_{\downarrow}$ is lost.
As $H_{\uparrow}$ and $H_{\downarrow}$ provide different sets of energy eigenvalues, the DOS spectra no longer match with each other for non-zero bias drop (Fig.~\ref{fig:f3} (b)). Moreover, we find a large gap across $E=0$, for $V=0$ case due to the binary nature of site energies, i.e, $-h,h,-h,h,.....$, and two sub-bands appear. This binary nature starts disappearing with a non-zero bias drop, and it vanishes in the limit of a large drop. 
 
Thus, the role of bias drop in the clean AFM chain in producing a mismatch among up and down spin channels is clear. Now, we will be 
focusing on how the bias drop affects different spin dependent transport phenomena. Figure~\ref{fig:f4} displays up and down spin 
transmission probabilities, as a function of energy $E$, under three different biased conditions. For the zero-biased condition, an 
exact overlap occurs between $T_{\uparrow}$ and $T_{\downarrow}$, whereas they are misaligned at non-zero biases, and these features are 
well understood from the previous analysis. From each sub figure it is found that the resonant peaks are obtained at some discrete 
energies. 
\begin{figure}[ht]
	\centering
	\begin{minipage}[b]{0.48\textwidth}
		\centering
		\resizebox{7.5cm}{5cm}{\includegraphics{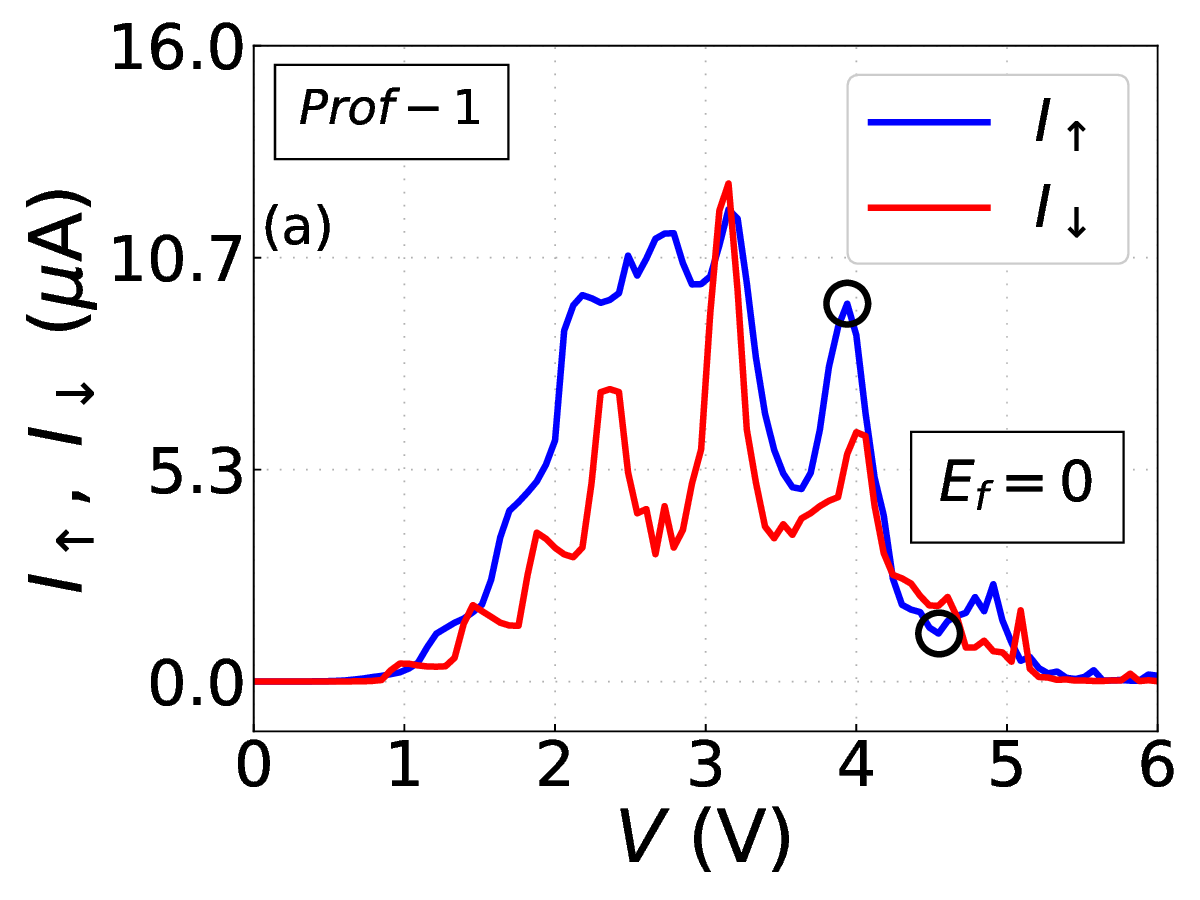}}
	\end{minipage}
	
	\begin{minipage}[b]{0.48\textwidth}
		\centering
		\resizebox{7.5cm}{5cm}{\includegraphics{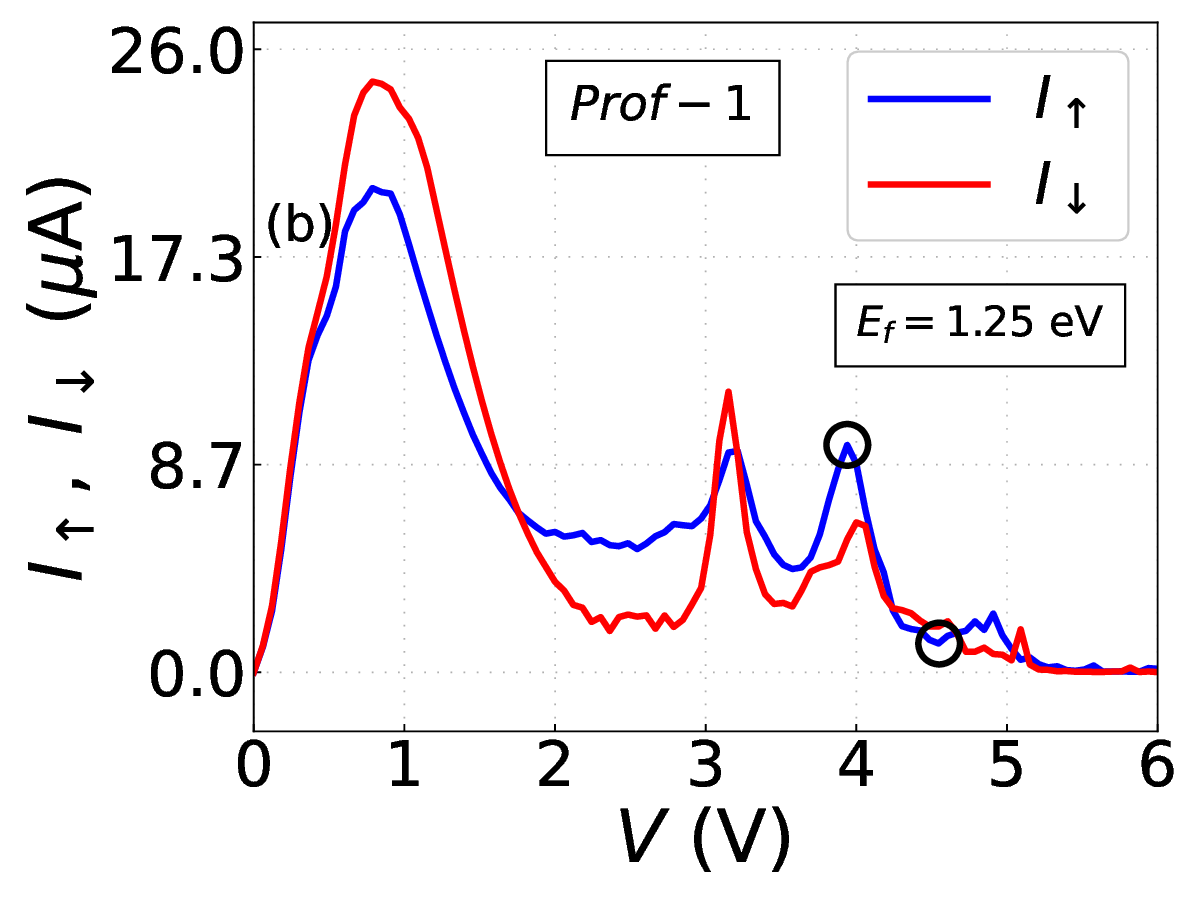}}
	\end{minipage}
	
\caption{(Color online). Up and down spin currents as a function of bias voltage, considering a linear bias drop along the clean AFM chain, at two different Fermi energies, where (a) $E_F=0$ and (b) $E_F=1.25\,$eV. In each sub-figure, two small black circles are drawn to indicate the 
peak and valley currents, in one of the NDR regions.}
	\label{fig:f5}
\end{figure}
These energies are associated with the eigenenergies of the AFM chain. 
A careful inspection reveals that the number of resonant peaks is not identical for the three values of $V$. It 
decreases with increasing $V$. This behavior is both interesting and physically significant. For a finite $V$, the site energies 
become non-uniform even when the system is impurity-free, thereby breaking the translational symmetry of the system. 
Such non-uniform site energies suppress inter-site 
electron hopping, leading to the localization of some eigenstates~\cite{loca} around individual lattice sites rather than being 
extended throughout the system. This phenomenon is referred to as bias-induced electronic localization, which is analogous to
electric-field-induced Wannier-Stark localization. The effect becomes more pronounced with increasing bias strength. With decreasing 
the number, the heights of the peaks also get shortened compared to unity due to enhanced electron scattering caused by the non-uniform site 
energies. Since localized states do not support electron transport, the total number of transmission peaks across the full energy window 
is reduced. For a sufficiently large $V$, all eigenstates become localized, and consequently, the transmission vanishes.

Once the transmission probability is found, the junction current can be easily computed. It is worth noting that for our junction setup, 
the transmission function is `voltage dependent', and thus, in order to calculate current, we need to evaluate the transmission probability 
at each and every voltage, unlike the usual junction setup where bias drop is considered only at the contact points.

Figure~\ref{fig:f5} shows the variation of up and down spin currents as a function of bias voltage, at two different Fermi energies 
where (a) $E_F=0$ and (b) $E_F=1.25\,$eV. The results are shown considering a linear bias drop (prof-1) along the clean AFM chain. Several
important features are obtained from the current-voltage spectra. First of all, a considerably large mismatch occurs between the two spin
current components for a wide bias window. This is due to the finite difference between up and down spin transmission profiles in the 
presence of a non-zero bias drop along the chain. For too weak biases, the mismatch becomes quite small, as expected. The crucial 
observation is that, each of the current components initially increases and then decreases with bias voltage.
\begin{figure}[h]
	\centering
	
	\begin{minipage}[b]{0.23\textwidth}
		\centering
		\resizebox{4cm}{3cm}{\includegraphics{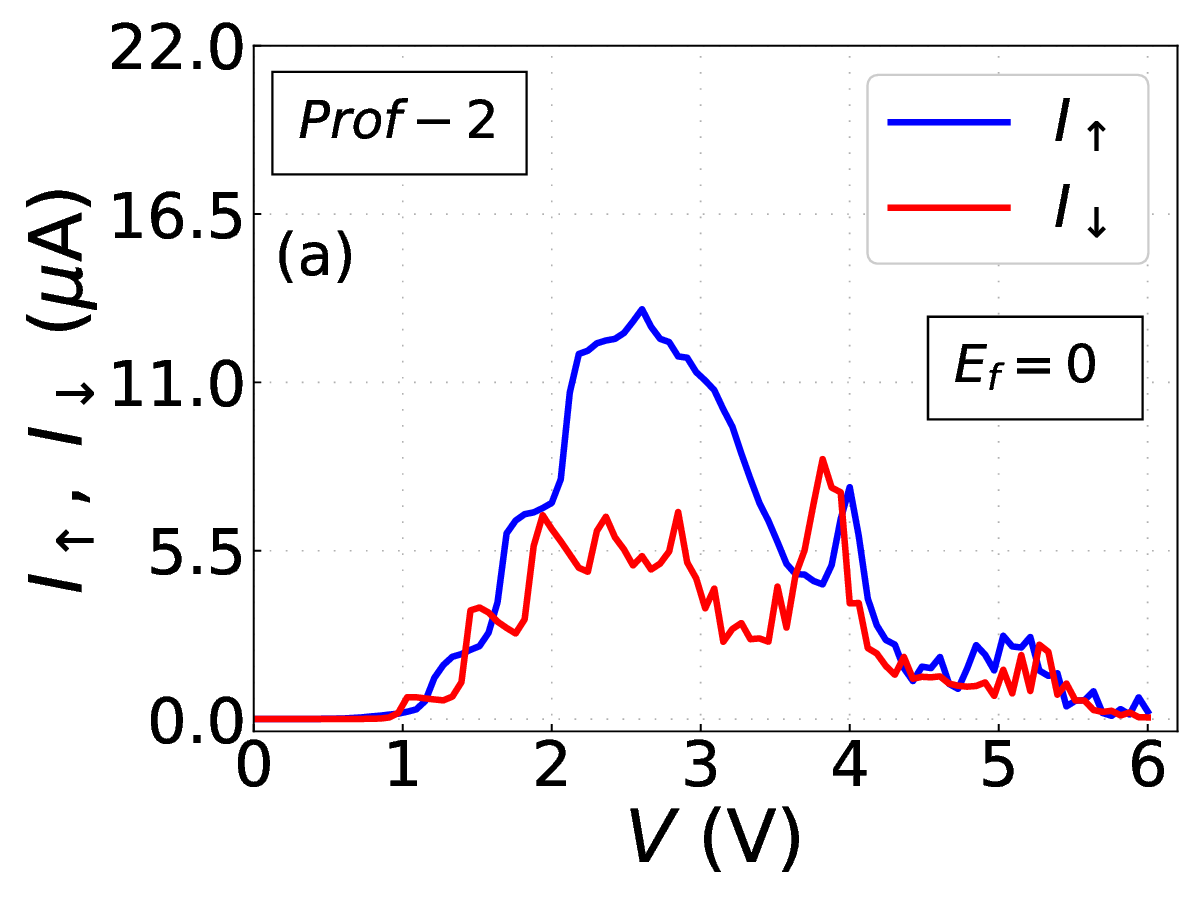}}
	\end{minipage}
	\hfill
	\begin{minipage}[b]{0.23\textwidth}
		\centering
		\resizebox{4cm}{3cm}{\includegraphics{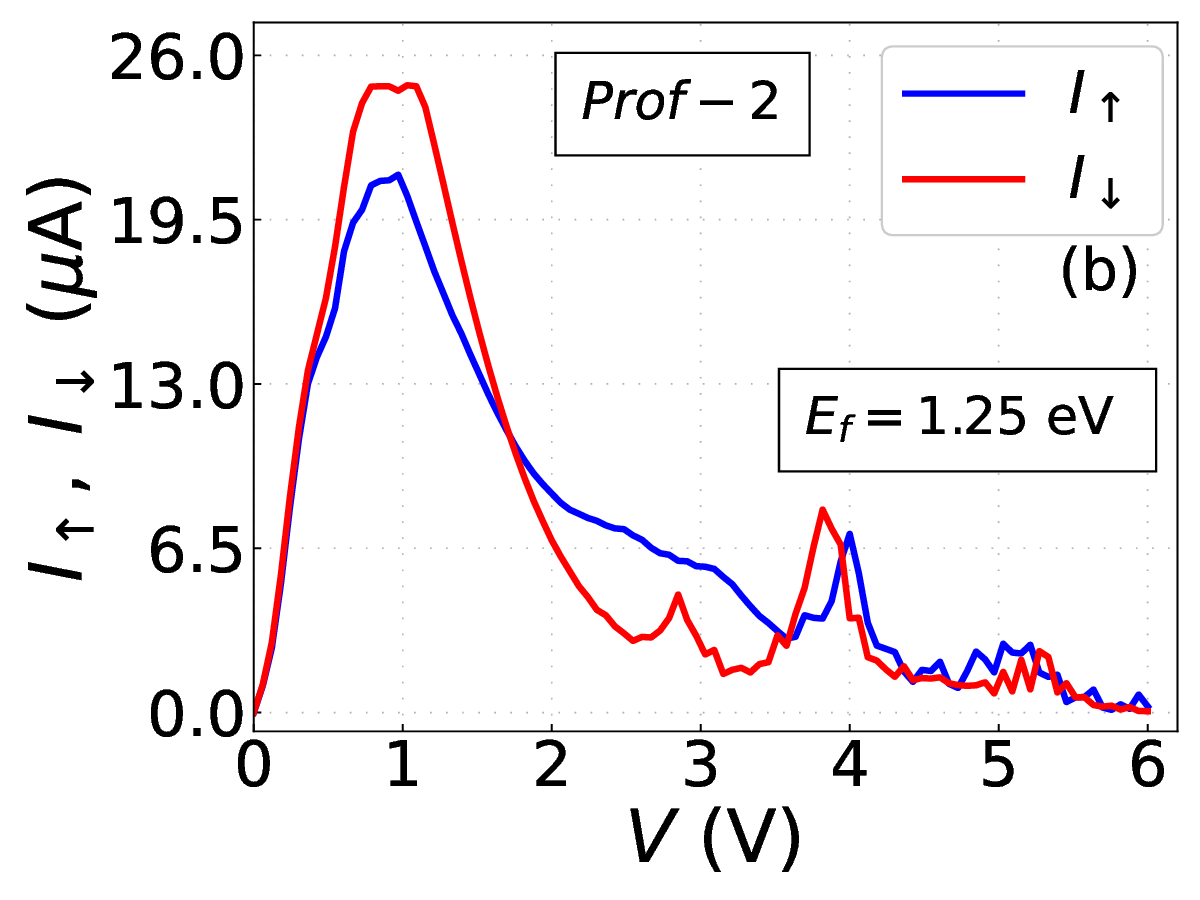}}
	\end{minipage}
	
	\vspace{0.5cm}
	
	\begin{minipage}[b]{0.23\textwidth}
		\centering
		\resizebox{4cm}{3cm}{\includegraphics{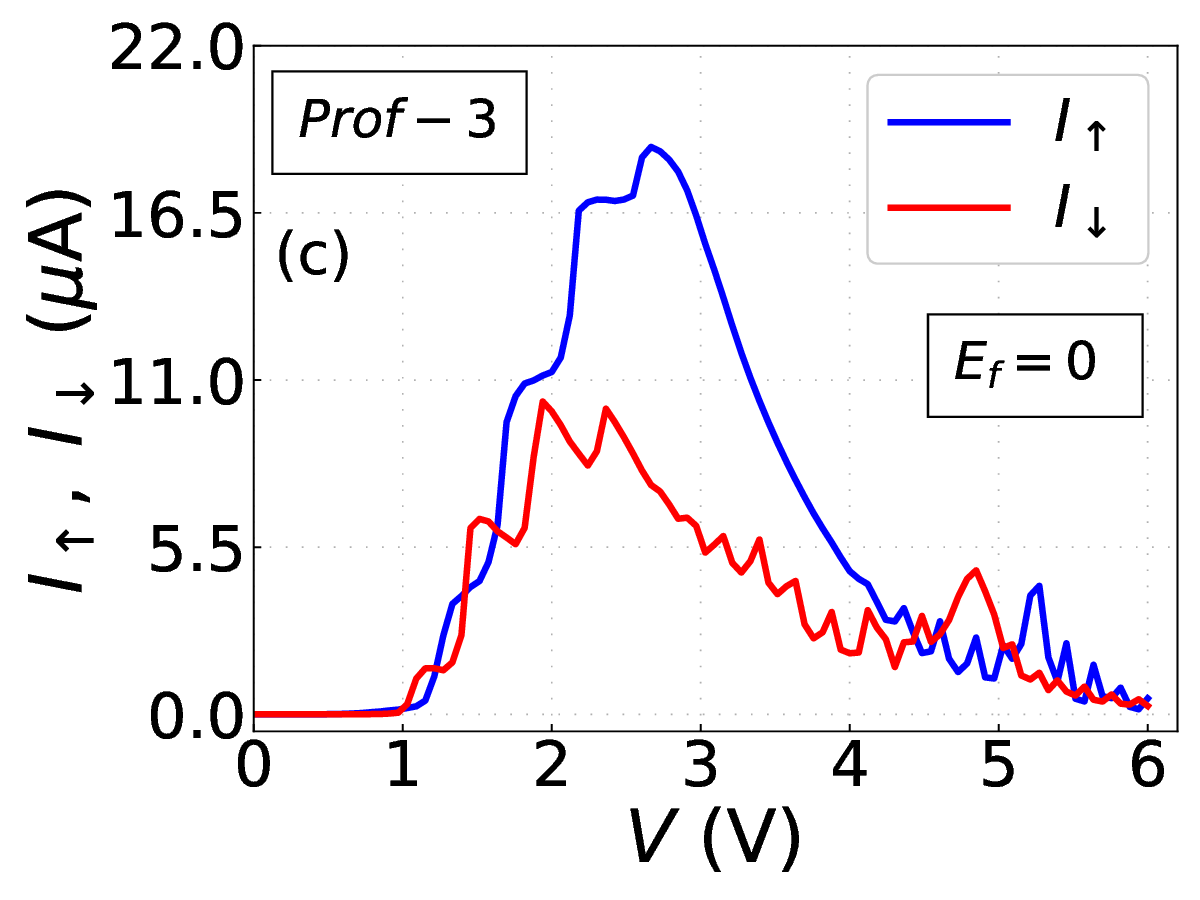}}
	\end{minipage}
	\hfill
	\begin{minipage}[b]{0.23\textwidth}
		\centering
		\resizebox{4cm}{3cm}{\includegraphics{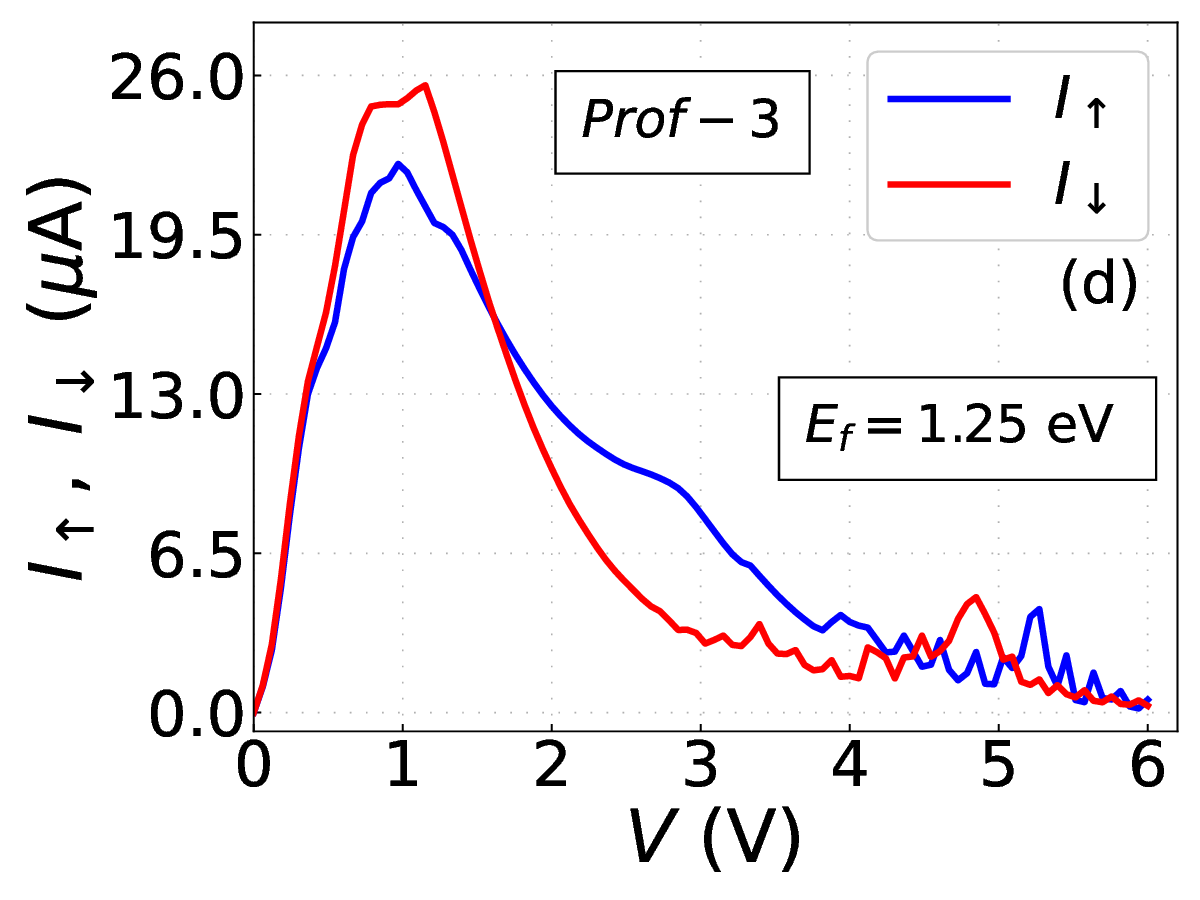}}
	\end{minipage}
\caption{(Color online). Up and down spin currents as a function of voltage in the presence of a non-linear bias drop along the clean
AFM chain,
where the upper and lowers rows are for the prof-2 and prof-3, respectively. In each case, the currents are shown for the two different
Fermi energies, like what are considered in Fig.~\ref{fig:f5}.}
\label{fig:f6}
\end{figure}
The reduction of the junction (transport) current with increasing voltage is unusual and is referred to as the negative differential 
resistance (NDR) effect. For a large enough voltage, the current almost vanishes. The increasing and decreasing nature may also be viewed 
at multiple times depending on the choice of the equilibrium Fermi energy $E_F$, that is visible both from Figs.~\ref{fig:f5}(a) and (b), 
but more prominent in Fig.~\ref{fig:f5}(b).

The underlying physics of the above mentioned phenomena is as follows. The junction current is obtained by integrating the transmission
profile, following the Landauer-B\"{u}ttiker prescription. At absolute zero temperature $(\Theta=0\,$K), the integration limit becomes $(E_F-eV/2)$ to $(E_F+eV/2)$, where $E_F$ corresponds to the equilibrium Fermi energy. For a chosen $E_F$, a non-zero current at any 
particular bias arises once any one of the transmission peaks appears within the energy window. 
\begin{figure}[ht]
	\centering
	\hspace*{-3.3cm}
	\begin{minipage}{0.42\columnwidth}
		\centering
		\resizebox{7.5cm}{5cm}{\includegraphics{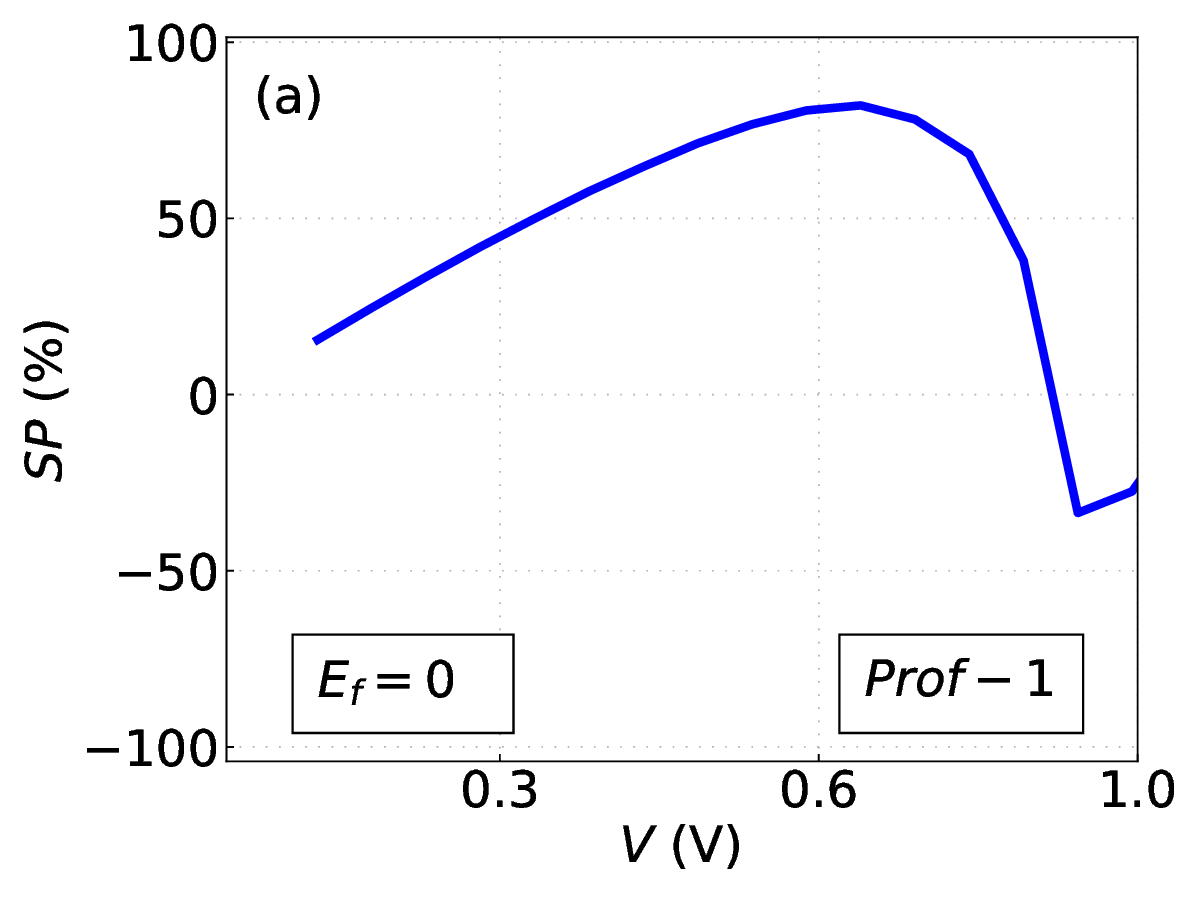}}
	\end{minipage}

	\hspace*{-3.3cm}
	\begin{minipage}{0.42\columnwidth}
		\centering
		\resizebox{7.5cm}{5cm}{\includegraphics{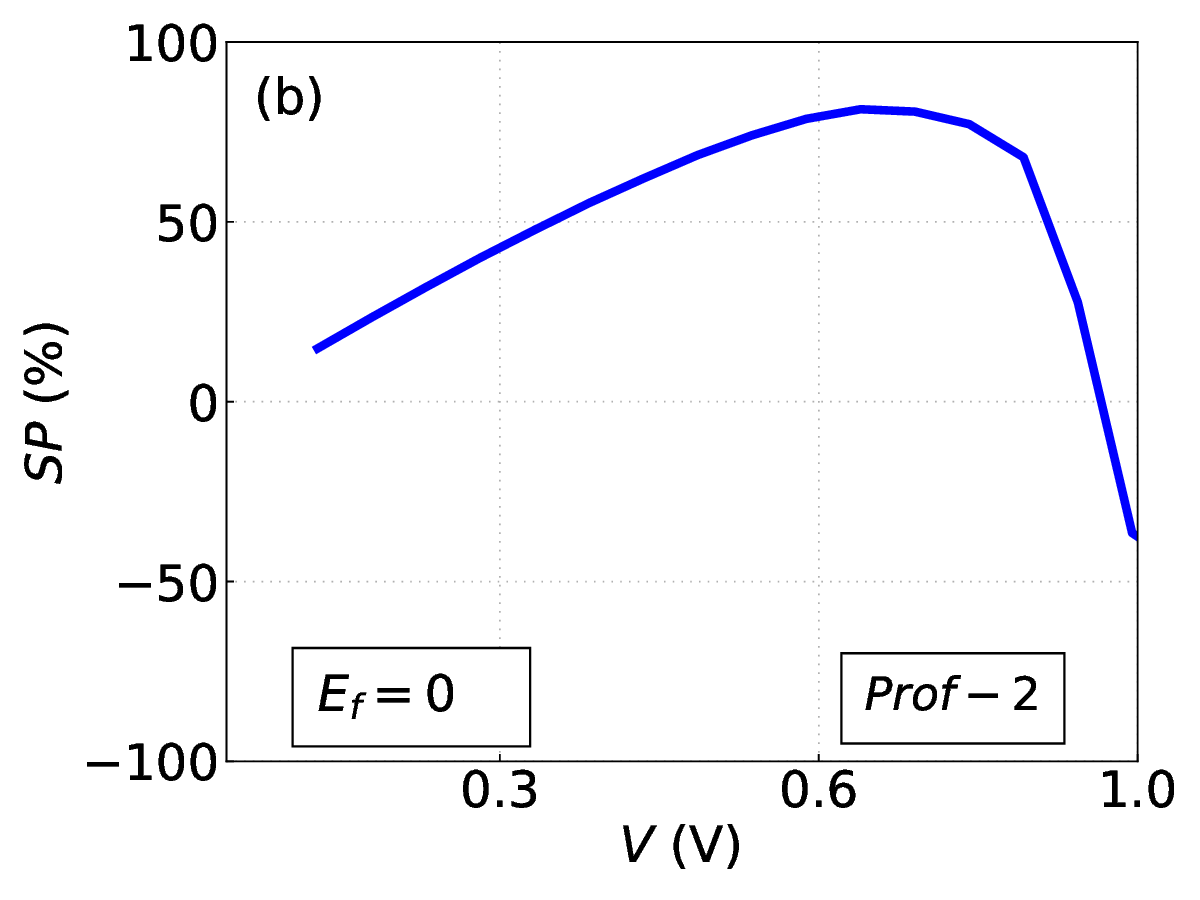}}
	\end{minipage}

	\hspace*{-3.3cm}
	\begin{minipage}{0.42\columnwidth}
		\centering
		\resizebox{7.5cm}{5cm}{\includegraphics{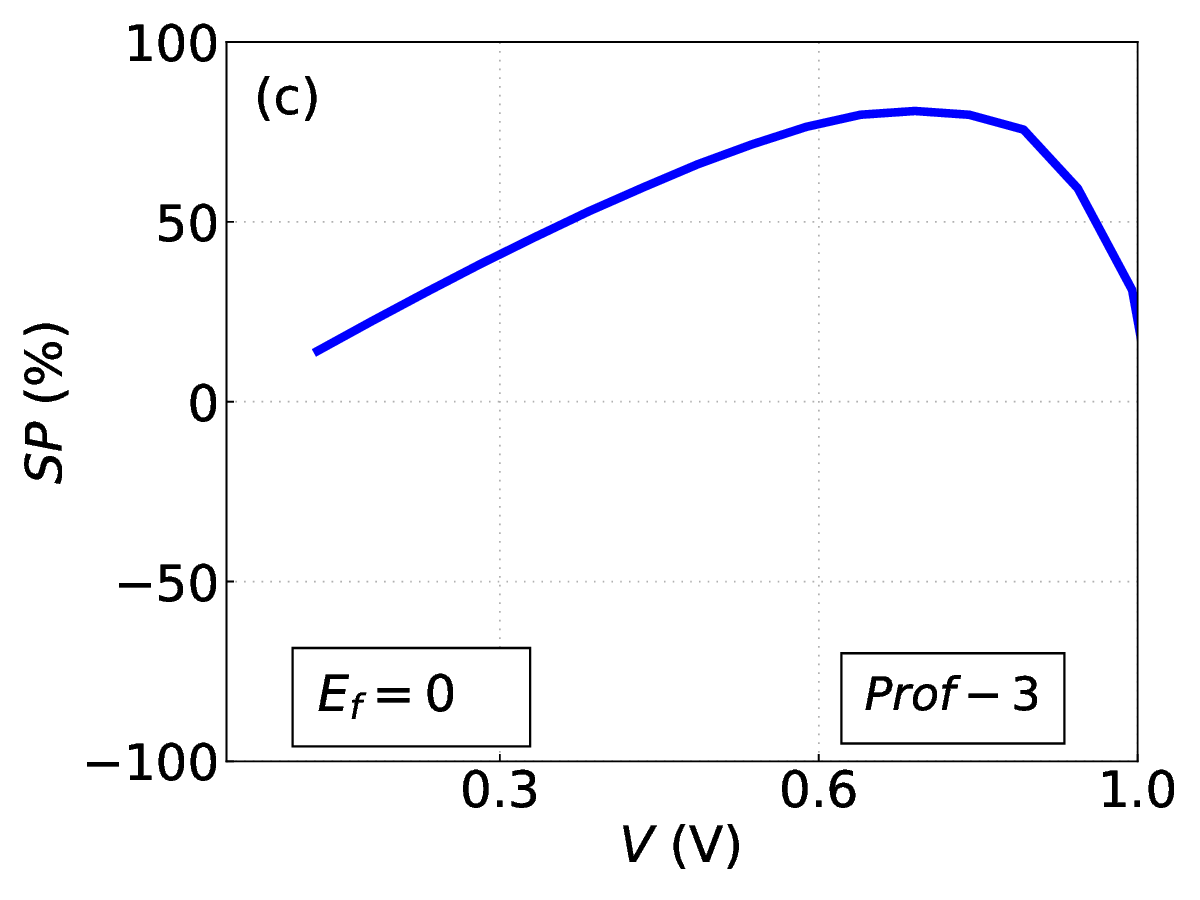}}
	\end{minipage}
	
\caption{(Color online). Variation of spin polarization, in the clean AFM chain, with bias voltage $V$ for the three different 
potential profiles, where (a) prof-1, (b) prof-2, and (c) prof-3, respectively. Here, the Fermi energy is set at zero.}
	\label{fig:f7}
\end{figure}
It is quite expected that more transmission peaks come within the window when the voltage is 
increased. But, unlike nanojunctions where the bias drop occurs only at the junction interfaces, for our present setup, it is not always
expected that more transmission peaks contribute to current with increased bias voltage, as the transmission probability is itself 
`voltage dependent'. The transmission peaks are associated with the energy eigenvalues of the chain placed between the contact electrodes. 
With changing the voltage $V$, the eigenspectrum of $H_{AFM}(V)$ gets changed, yielding a voltage-dependent transmission probability. At 
low biases, when eigenenergies are not that much modified, more peaks are accommodated, resulting in an enhanced current with $V$. But, 
at higher biases, some resonant transmission peaks go out the window, and some peaks are shortened and narrowed due to non-uniform effective
site energies. As a result, the current gets reduced with voltage, providing the NDR phenomenon. As the transmission peaks are redistributed
whenever the voltage gets changed, the appearance of the NDR phenomenon at multiple regions with the specific choice of $E_F$ is expected. 
In the end, when all the states are localized at too high bias, the currents drop to zero (Fig.~\ref{fig:f5}).

As pointed out earlier, the performance of the NDR phenomenon is quantified by the peak-to-valley current ratio (PVCR). From each sub-figure 
of Fig.~\ref{fig:f5}, we choose one region among multiple NDR regions and mark the peak and valley currents with small black circles. The 
PVCR values are $7.86$ and $7.9$ respectively which are relatively larger compared to most of the reported results available in the 
literature~\cite{PVCR1,PVCR2,PVCR4,PVCR3}. To inspect how the above-discussed results are sensitive to the other choices of potential 
profiles, 
\begin{figure}[h]
	\centering
	\hspace*{-3.3cm}
	\begin{minipage}{0.42\columnwidth}
		\centering
		\resizebox{7.5cm}{5cm}{\includegraphics{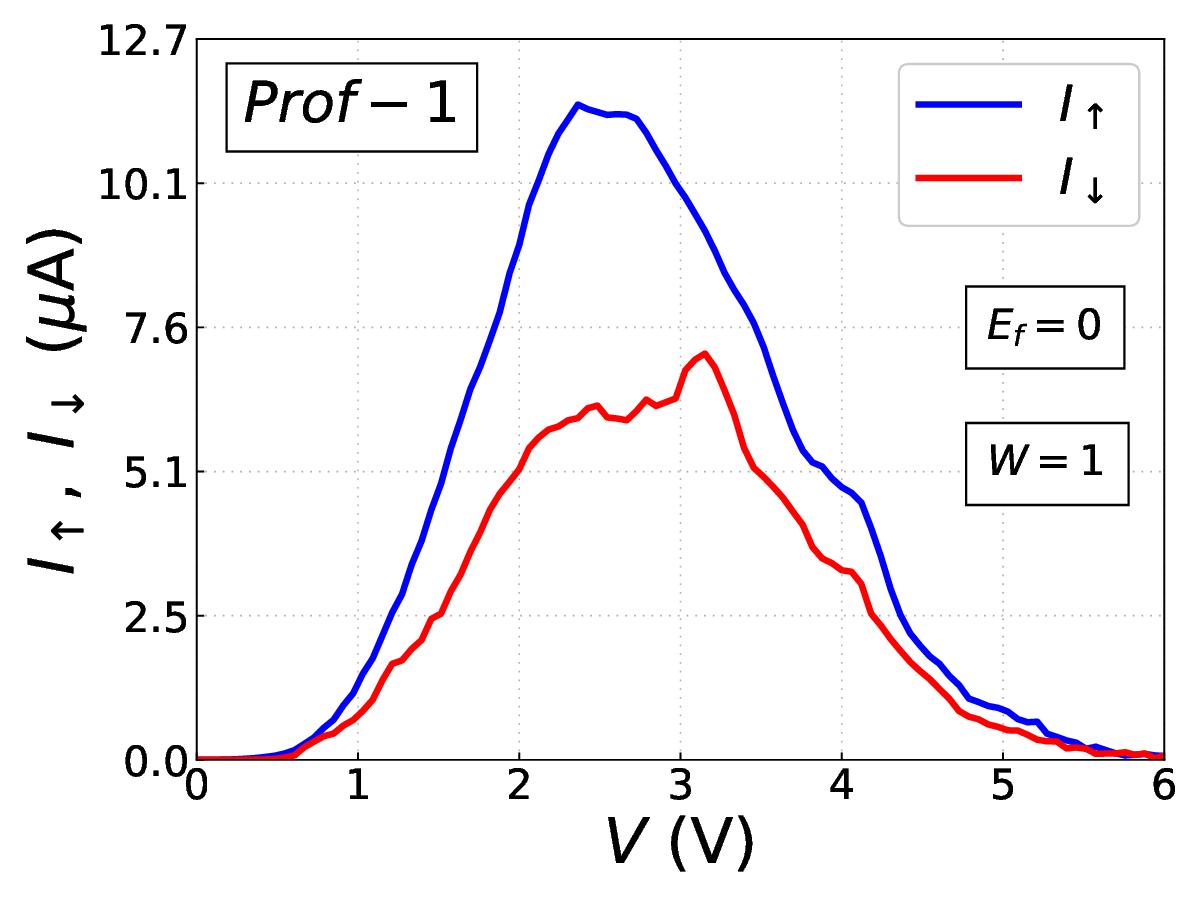}}
	\end{minipage}          
	\caption{(Color online). Effect of disorder: Up and down spin currents as a function of bias voltage in the presence of random 
		(uncorrelated) disorder, with disorder strength $W=1$, considering a linear bias drop along the AFM chain. The Fermi energy is set at zero.}
	\label{dis}
\end{figure}
in Fig.~\ref{fig:f6} we present the
spin-dependent currents as a function of bias voltage for two non-linear potential profiles, prof-2 and prof-3. In each case, we compute 
the currents for two distinct Fermi energies. The results are quite similar to those illustrated in Fig.~\ref{fig:f5} for the linear bias 
drop (prof-1). A significant mismatch between the two spin currents is observed, along with the NDR phenomenon appearing in multiple 
bias regions. A careful comparison of the current magnitudes from Figs.~\ref{fig:f5} and \ref{fig:f6} reveals that the current magnitude
increases with the flatness of the potential profile along the AFM chain. The maximum current amplitudes for the three profiles (prof-1 
to prof-3) are $20.21$, $21.28$, and $22.39$, respectively. This enhancement is directly associated with the decreasing non-uniformity of 
the effective site energies as the potential profile becomes flatter. The highest maximum current is expected when the bias drop occurs 
only at the junction interfaces (not shown here).   

The large mismatch between the up and down spin currents indicates a favorable spin polarization (SP). To quantify this, in 
Fig.~\ref{fig:f7} we show the variation of SP as a function of bias voltage for the chosen potential profiles, one linear and two non-linear. 
In all cases, the overall pattern remains quite similar, the degree of SP initially increases with voltage, reaches a maximum, and then
gradually decreases. At its peak, SP reaches $\sim 90\%$, which is highly significant. The rate of decrease in SP beyond the critical voltage,
where SP attains its maximum, becomes more pronounced with the increasing steepness of the bias drop along the system. 
\begin{figure}[h]
	\centering
	\begin{minipage}[b]{0.42\textwidth}
		\centering
		\resizebox{7.5cm}{5cm}{\includegraphics{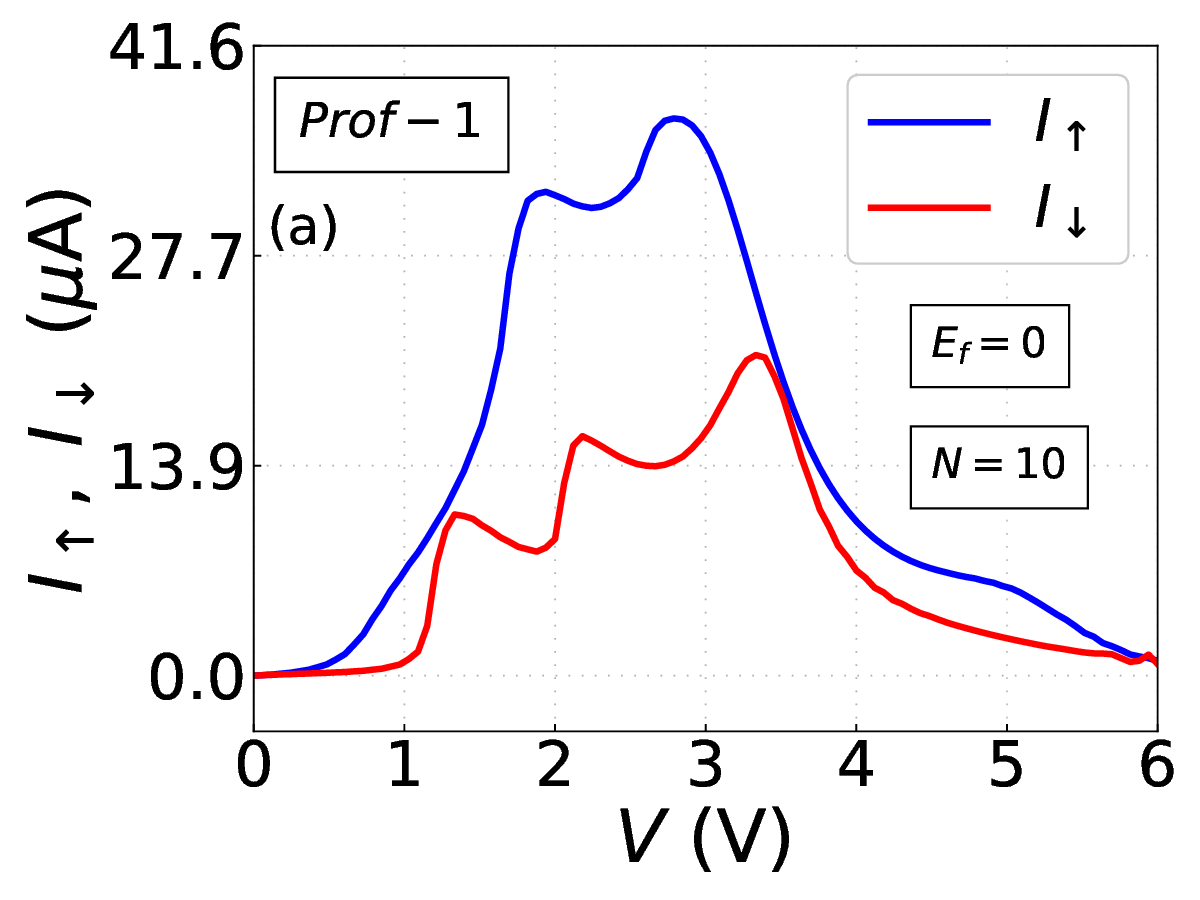}}
	\end{minipage}
	\begin{minipage}[b]{0.42\textwidth}
		\centering
		\resizebox{7.5cm}{5cm}{\includegraphics{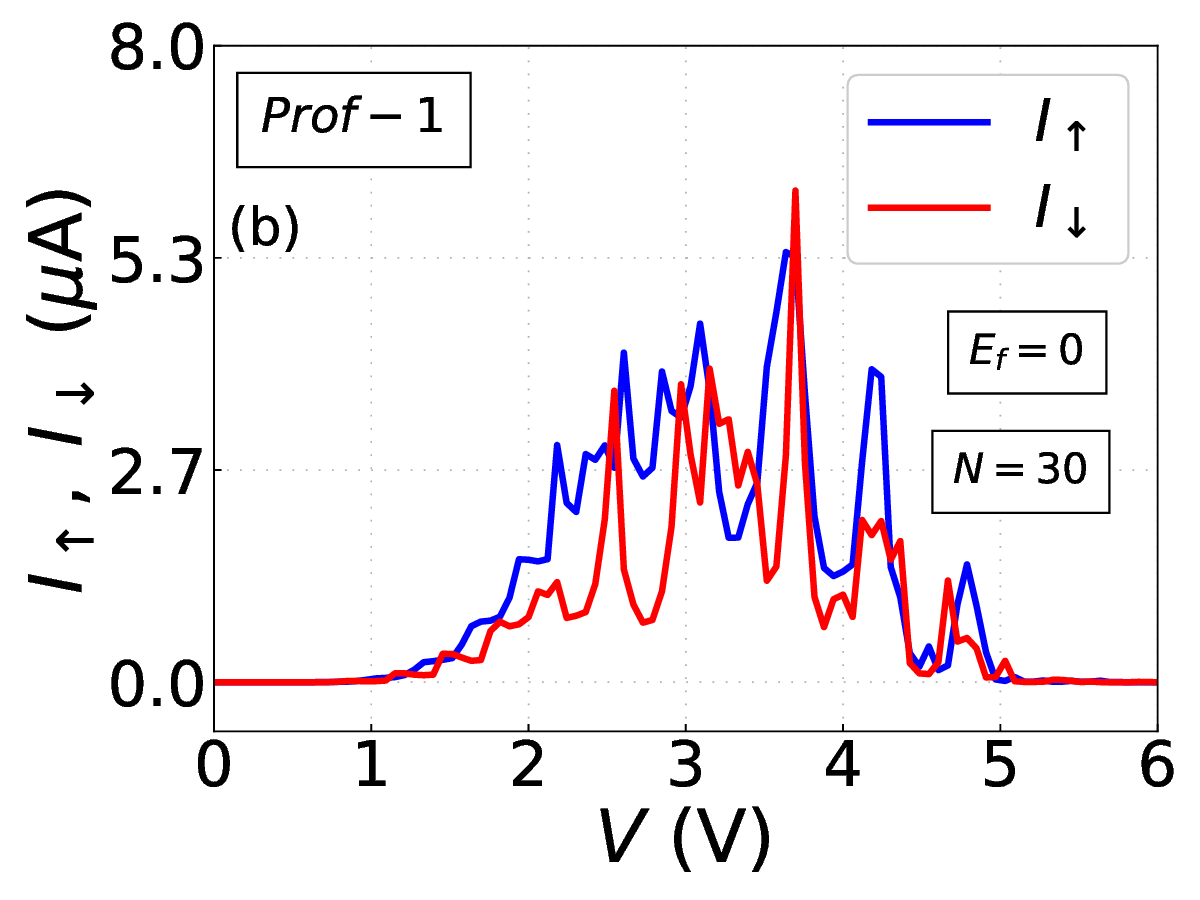}}
	\end{minipage}       
	\caption{(Color online). Effect of system size $N$: Up and down spin currents as a function of voltage for the disorder-free AFM chain 
		($W=0$) considering two different chain lengths, where (a) $N=10$ and (b) $N=30$. The bias drop along is the chain is considered following 
		the prof-1, and the results are worked out at $E_F=0$.} 
	\label{size}
\end{figure}
Although the asymmetry
between the up and down spin sub-Hamiltonians increases with potential steepness, which in principle enhances the mismatch between the two 
spin channels, it simultaneously increases the likelihood of localization of the energy eigenstates. The combined effect of these two 
competing factors is reflected in the SP-$V$ characteristics.  

To make the proposed quantum system more realistic, we include the effect of substitutional disorder in the AFM chain, by choosing the site
energies, $\epsilon_n^{V=0}$, randomly from a `Box' distribution function of width $W$. For the clean AFM chain, the disorder strength $W=0$.
We specifically want to check whether the results discussed earlier, for the clean system, are still valid in the presence of disorder. 
As illustrative example, in Fig.~\ref{dis} we show the variations of two different spin current components as a function of bias voltage,
considering the disorder strength $W=1$, and setting the Fermi energy $E_F=0$. Since the site energies are uncorrelated, we take a large 
number of distinct disordered configurations ($50$ in total) and compute the configuration-averaged results. Looking at the red and blue
curves, it can be emphasized that all the physical phenomena viz, the appearance of a large mismatch between the up and down spin currents and
the reduction of current with voltage beyond a critical value, remain the same. 
\begin{figure}[h]
	\centering
	\begin{minipage}[b]{0.4\textwidth}
		\centering
		\resizebox{7.5cm}{5cm}{\includegraphics{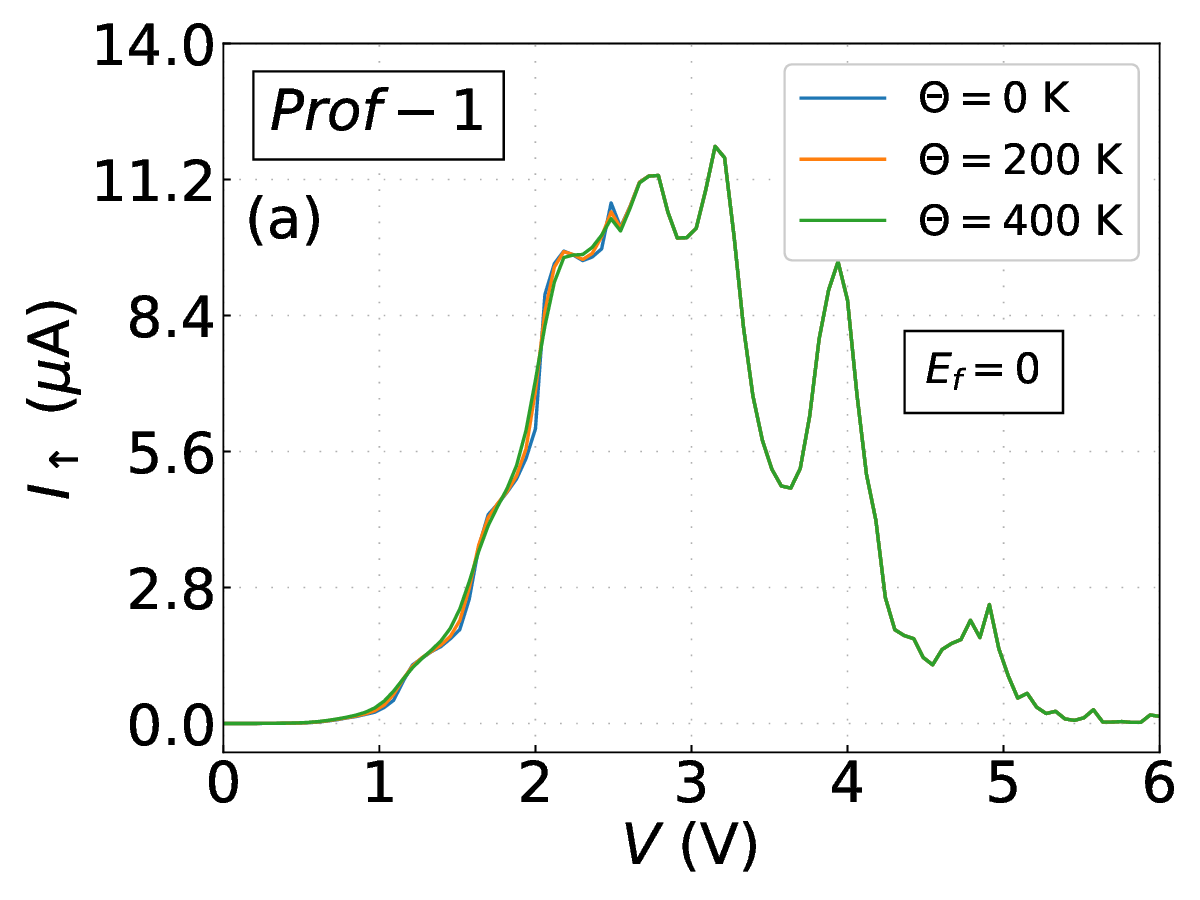}}
	\end{minipage}
	
	\begin{minipage}[b]{0.4\textwidth}
		\centering
		\resizebox{7.5cm}{5cm}{\includegraphics{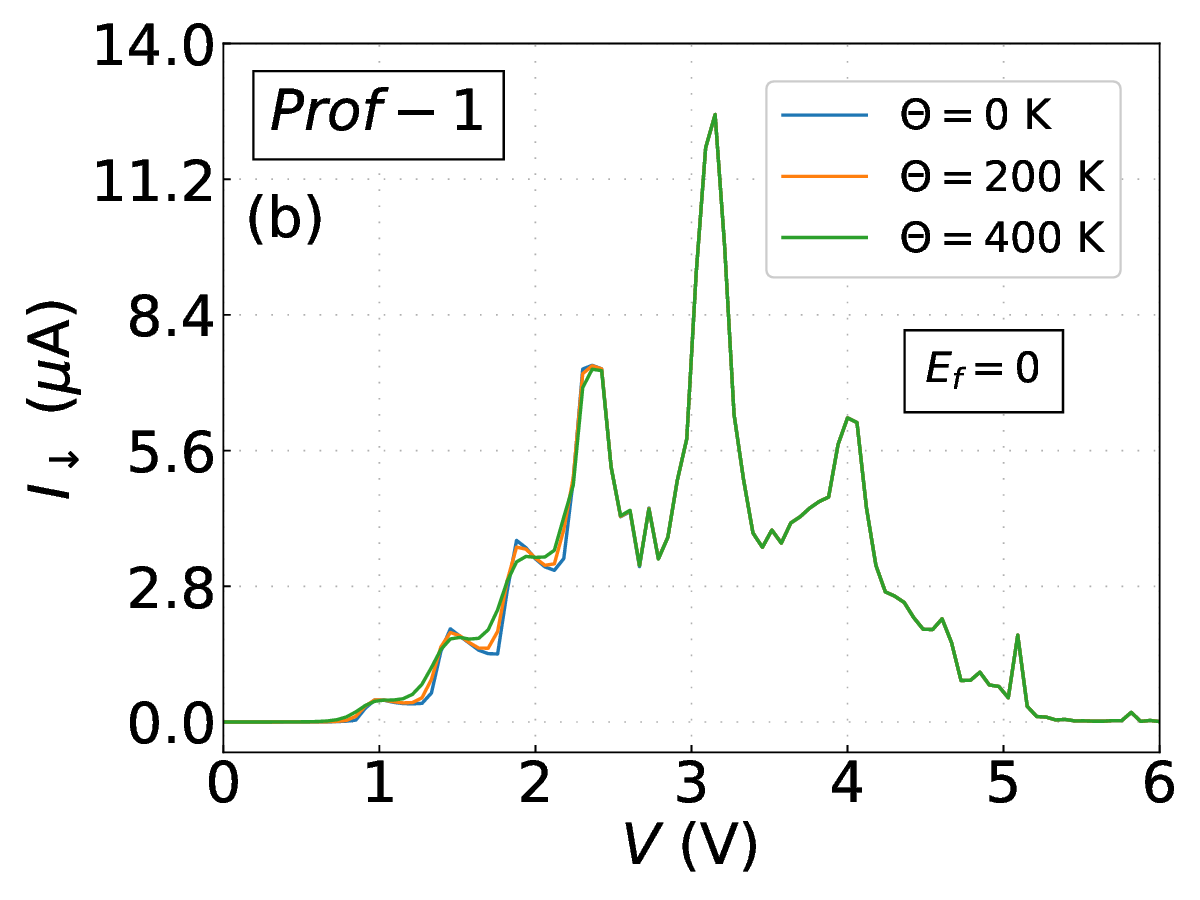}}
	\end{minipage}
	
\caption{(Color online). Effect of temperature: Up and down spin currents as a function voltage, shown in (a) and (b) respectively, for
a clean AFM chain with $N=20$, at two non-zero temperatures. The result of zero temperature is also superimposed in each spectrum. The 
linear bias drop, following the prof-1, is taken into account.}
\label{temp}
\end{figure}
In addition, it is relevant to point out that, in the presence
of disorder, the current magnitude decreases compared to what are observed for the disorder-free cases, and this is quite obvious. However, 
all the essential physical features persist until the disorder strength becomes sufficiently large to localize all the electronic states.  
 
In the same footing, to examine whether the physical phenomena persist for other system sizes, we consider two different chain lengths 
and present the results in Fig.~\ref{size}. All the basic features remain the same as before. However, a careful inspection reveals some
additional characteristics. For instance, with increasing chain length, the degree of misalignment between the up and down spin currents
decreases. This occurs because, as $N$ increases, a larger number of spin-dependent energy channels become available within a given voltage
window, thereby reducing their difference and resulting in a smaller mismatch. The NDR phenomenon is also observed in multiple voltage
regions for longer chains compared to shorter ones. Thus, the likelihood of observing NDR behavior increases with system size. It is also 
worth noting that very long chains cannot be considered, since in that case, the electronic states may become localized due to the bias
drop along the chain. Therefore, a moderate chain length is highly recommended.

The results analyzed so far have been obtained at absolute zero temperature ($\Theta = 0\,$K).
\begin{figure}[h]
	\centering
	\begin{minipage}[b]{0.23\textwidth}
		\centering
		\resizebox{4.5cm}{3.5cm}{\includegraphics{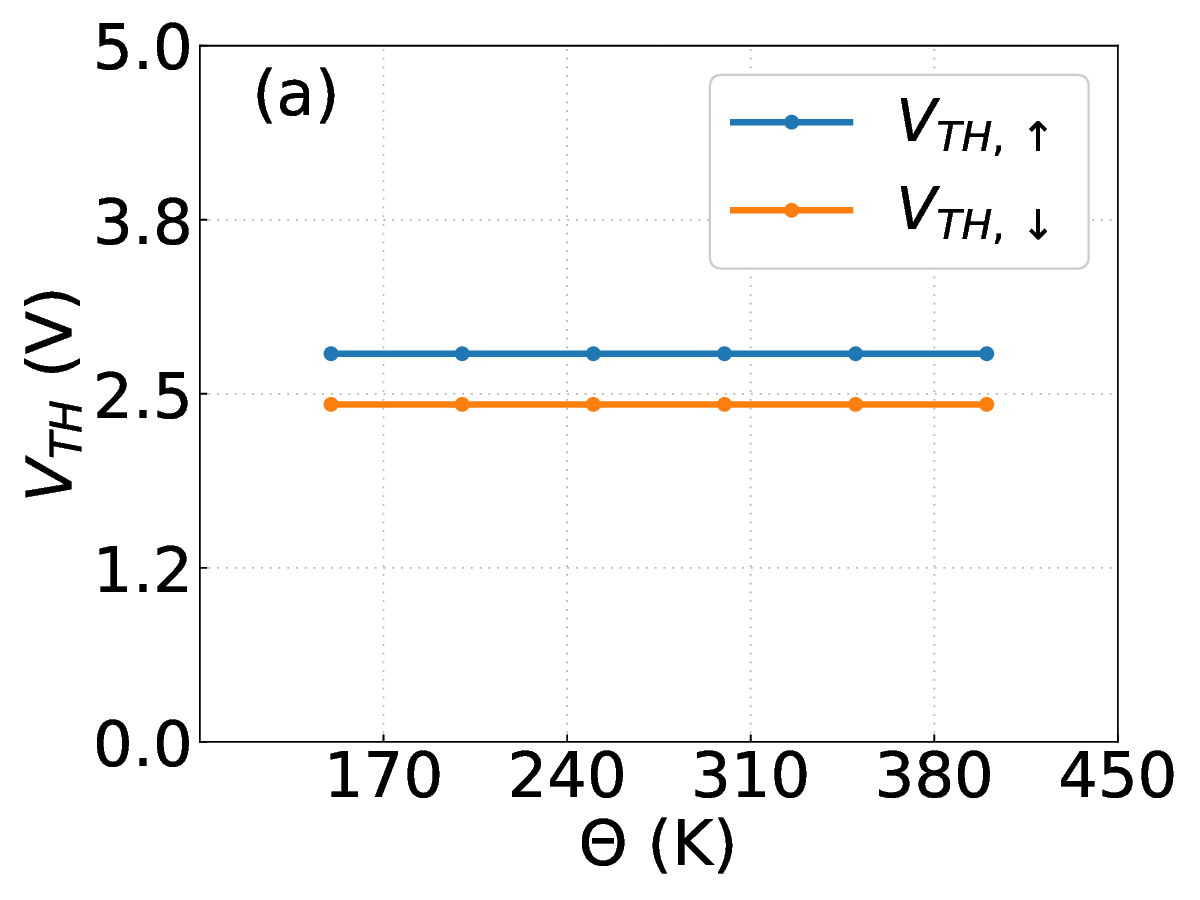}}
	\end{minipage}
	\hfill
	\begin{minipage}[b]{0.23\textwidth}
		\centering
		\resizebox{4.5cm}{3.5cm}{\includegraphics{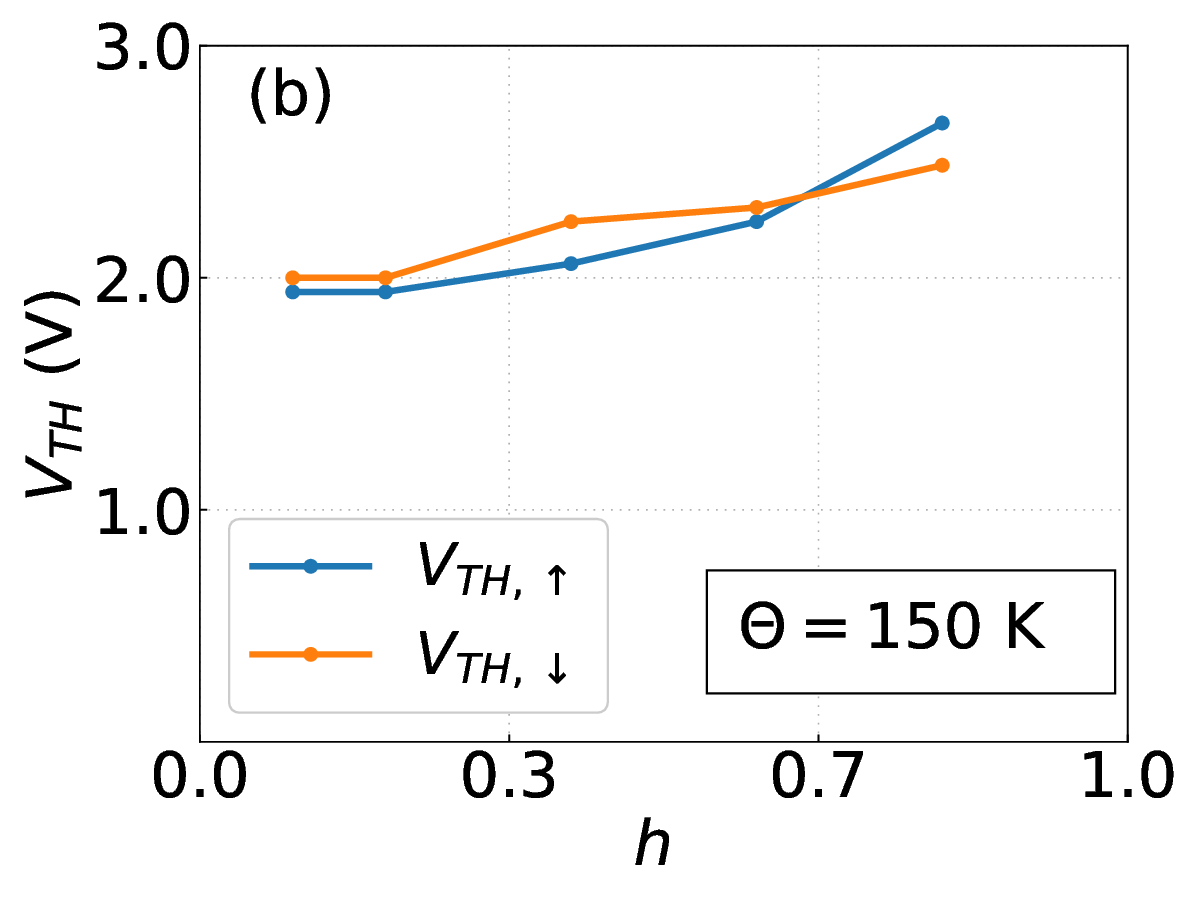}}
	\end{minipage}
	
	\vspace{0.5cm}
	
	\begin{minipage}[b]{0.25\textwidth}
		\centering
		\resizebox{5cm}{4cm}{\includegraphics{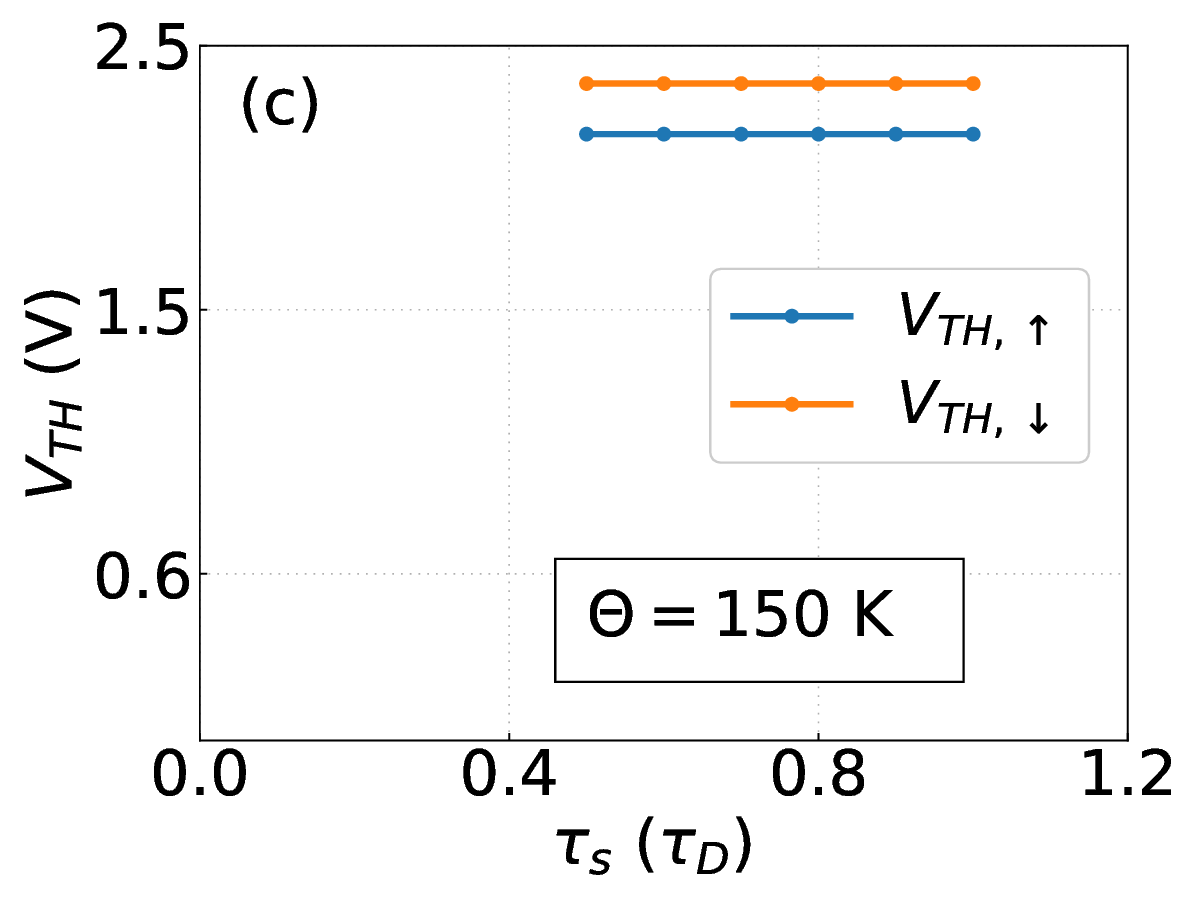}}
	\end{minipage}
	
\caption{(Color online). Variation of threshold voltage $V_{TH}$, in the clean AF chain, with (a) temperature ($\Theta$), 
(b) spin-dependent scattering factor ($h$), and (c) chain-to-electrode coupling strengths ($\tau_S, \tau_D$). We assume 
$\tau_S=\tau_D$. For (b) and (c), the temperature is fixed at $\Theta=150\,$K. The other parameters are: $N=20$ and $E_F=0$.}
	\label{figvth}
\end{figure}
To make the study more realistic and complete, we now examine the effect of finite temperature and discuss various related aspects. At non-zero temperatures, the influence of thermal broadening must be incorporated into the current calculation through the term ($f_S-f_D$), with the integration performed over the entire allowed energy window. Figure~\ref{temp} illustrates the temperature dependence of the up- and down-spin currents in a clean AFM chain of length $N=20$. Two finite temperatures, $200\,$K and $400\,$K, are considered, represented by the red and green curves, respectively. For comparison, the zero-temperature result is also shown (blue curve). Interestingly, the temperature has only a marginal effect on the currents. For both spin channels, the currents remain almost comparable, becoming nearly indistinguishable at higher bias voltages.
Although thermal broadening is present, its effect is quite weak due to the small system size. For a short chain, the average level spacing is relatively large, leading to well-separated resonant transmission peaks, and thus, thermal broadening cannot significantly alter the transport behavior. This is indeed a favorable outcome. The notable mismatch between the two spin currents and the appearance of the NDR phenomenon persist even at finite temperatures, similar to the zero-temperature case.

Figure~\ref{figvth} illustrates the dependence of the threshold bias voltage ($V_{TH}$) on various tight-binding parameters, such as 
temperature ($\Theta$), spin-dependent scattering strength ($h$), and the chain-to-electrode coupling strengths ($\tau_S$ and $\tau_D$). 
The threshold voltage is determined by identifying the bias at which the first NDR feature emerges. The results show that $V_{TH}$ is 
largely insensitive to these parameters, with only a slight variation observed for different values of $h$. These findings clearly indicate 
the robustness of the spin-specific NDR phenomenon in a {\em clean} AFM chain, where the applied bias alone is sufficient to break the 
symmetry between the up and down spin sub-Hamiltonians.  

\vskip 0.2cm
\noindent
\textbf{Experimental Perspective}: To ensure that our theoretical framework remains experimentally verifiable, it is important to explore
possible realizations at the nanoscale. Scanning tunneling microscopy (STM)-based studies have demonstrated that atomic-scale 
antiferromagnetic (AFM) chains composed of Fe atoms can be precisely constructed and manipulated on a $Cu_2N/Cu(100)$ surface at 
low temperatures~\cite{d4}. Similar chains can also be engineered using Mn atoms as an alternative to Fe. In a recent
experiment, Su {\em et al.} successfully fabricated an antiferromagnetic spin-$1/2$ Heisenberg chain through a combined on-surface synthesis 
and reduction technique. In their approach, closed-shell oligomers were transformed into spin chains by controlled STM-tip manipulation 
followed by hydrogen treatment~\cite{ex1}.

In our analysis, the currents are shown over a wide range of bias voltages to capture the full picture, namely the 
enhancement, reduction, and eventually vanishing nature. From an experimental perspective, it might seem that the voltage range is quite 
high, but this issue can easily be managed simply by adjusting the NNH strength of the AFM chain. Then, the full picture can be captured 
within a narrow voltage window, keeping all the physics unchanged.

When considering the experimental realization of our system, particular attention must be paid to the system size. Since 
we are dealing with spin-dependent transport phenomena, the chain length should be chosen such that it remains within the spin coherence 
length, $L_{\phi}$. The reported value of $L_{\phi}$ is of the order of $250\,$nm~\cite{sch1,sch2}, which implies that our results can be 
safely validated for a wide range of chain lengths.

\section{Closing Remarks and Outlook}

We have proposed and analyzed a simple yet robust mechanism for achieving spin-selective electron transport in a magnetic nanojunction 
with zero net magnetization. Unlike conventional approaches that rely on intrinsic spin-orbit coupling or magnetic asymmetry, our method
introduces a bias drop along the system to break the symmetry between the up- and down-spin sub-Hamiltonians. Using a tight-binding model 
of an antiferromagnetic (AFM) chain with antiparallel local moments, spin-dependent transmission probabilities are calculated via 
wave-guide theory, and the corresponding spin currents are evaluated using the Landauer-B\"{u}ttiker formalism.

Our results reveal highly spin-polarized currents across a wide bias range, even in the absence of net magnetization. Additionally, the
bias-dependent transport characteristics exhibit clear negative differential resistance (NDR) features, which persist for different 
potential profiles, both linear and non-linear, confirming the generality of the effect. The phenomena are found to be robust against 
variations in temperature, electrode coupling, and other tight-binding parameters, with only minor dependence on the spin-dependent 
scattering strength.

The proposed mechanism is experimentally feasible, as similar AFM chains, such as Fe or Mn atom chains on a $Cu_2N/Cu(100)$ surface 
or on-surface synthesized spin-$1/2$ chains, can already be fabricated and manipulated using scanning-probe techniques. Our findings 
offer a promising foundation for designing next-generation spintronic devices, such as bias-tunable spin filters and NDR-based 
functional elements, operating without any net magnetic moment.

For the benefit of readers, here we would like to point out that the present study involves some important 
approximations. The potential profile along a biased chain is essentially governed by electron screening and charge distribution. 
In our calculation, we have  considered three different potential profiles as illustrative examples, following earlier studies. These
profiles are spin independent and are identical both for up- and down-spin electrons. An accurate determination of such a potential 
profile can be obtained from a self-consistent Poisson-Schr\"{o}dinger treatment, which has not been adopted in this work and may be 
addressed in our future studies. While a self-consistent procedure may quantitatively renormalize the bias profile, the qualitative 
features of all the results presented here are expected to remain unchanged for the chosen bias profiles.
We have also not included the electron-electron interaction term in the Hamiltonian, and the model has been described within a 
single-particle framework. This approximation arises for two reasons. First, a one-dimensional half-filled Hubbard chain naturally 
exhibits antiferromagnetic ordering, and this ordering becomes more pronounced with increasing Hubbard i.e., the electron-electron 
Coulomb repulsion strength. Second, our focus is to explore the symmetry-breaking mechanism in a clean AFM chain. If a Hubbard term 
were included in the Hamiltonian, a stronger AFM ordering would emerge, which would in turn lead to a more pronounced separation 
between the spin channels in presence of a bias drop along the chain. That effect has been discussed in an indirect manner, as we 
have analyzed the results by varying the spin-moment scattering factor ($h$), which is related to the strength of the local magnetic 
moments at different lattice sites and is governed by the Hubbard interaction strength. In another approximation, we have neglected 
the effect of spin-orbit (SO) coupling in the present study. This is a reasonable assumption, as, on the one hand, the SO coupling is 
much weaker than the spin-moment interaction strength ($h$), and, on the other hand, in a one-dimensional two-terminal junction setup, 
SO coupling alone does not break Kramer's degeneracy and therefore does not lead to spin-channel separation. An elaborate discussion 
on this issue can be found in Ref.~\cite{cuv}.

\end{document}